\documentclass[twocolumn,prl,english,amsmath,amssymb,superscriptaddress,aps,longbibliography]{revtex4-1}
\usepackage[latin1,utf8]{inputenc}
\usepackage[T1]{fontenc}
\usepackage{amsmath}
\usepackage{amssymb}
\usepackage{amsthm}
\usepackage{amsfonts}
\usepackage{listings}
\usepackage{physics}
\usepackage[normalem]{ulem}
\usepackage{enumerate}
\usepackage{latexsym}
\usepackage{psfrag}
\usepackage{bm}
\usepackage{graphicx}
\usepackage{blkarray}
\usepackage{array}
\usepackage{color}
\usepackage[normalem]{ulem}

\usepackage[dvipsnames]{xcolor}
\definecolor{goodgreen}{rgb}{0.1,0.5,0}
\definecolor{goodred}{rgb}{0.7,0,0}
\usepackage[colorlinks,urlcolor=goodgreen,citecolor=blue,linkcolor=goodred]{hyperref}

\begin{document}
\title{Long-Range Propagation and Interference of $d$-wave Superconducting Pairs in Graphene}
\author{D. Perconte}
\affiliation{Unit\'e Mixte de Physique, CNRS, Thales, Universit\'e Paris-Sud, Universit\'e Paris-Saclay, 91767, Palaiseau, France}
\affiliation{Laboratorio de Bajas Temperaturas y Altos Campos Magn\'eticos, Departamento de F\'isica de la Materia Condensada, Instituto Nicol\'as Cabrera and Condensed Matter Physics Center (IFIMAC), Universidad Aut\'onoma de Madrid, E-28049 Madrid, Spain}
\author{K. Seurre}
\affiliation{Unit\'e Mixte de Physique, CNRS, Thales, Universit\'e Paris-Sud, Universit\'e Paris-Saclay, 91767, Palaiseau, France}
\author{V. Humbert} 
\affiliation{Unit\'e Mixte de Physique, CNRS, Thales, Universit\'e Paris-Sud, Universit\'e Paris-Saclay, 91767, Palaiseau, France}
\author{C. Ulysse}
\affiliation{Centre for Nanoscience and Nanotechnology, CNRS, Universit\'e Paris-Sud/Universit\'e Paris-Saclay, Boulevard Thomas Gobert, Palaiseau, France}
\author{A. Sander} 
\affiliation{Unit\'e Mixte de Physique, CNRS, Thales, Universit\'e Paris-Sud, Universit\'e Paris-Saclay, 91767, Palaiseau, France}
\author{J. Trastoy} 
\affiliation{Unit\'e Mixte de Physique, CNRS, Thales, Universit\'e Paris-Sud, Universit\'e Paris-Saclay, 91767, Palaiseau, France}
\author{V. Zatko} 
\affiliation{Unit\'e Mixte de Physique, CNRS, Thales, Universit\'e Paris-Sud, Universit\'e Paris-Saclay, 91767, Palaiseau, France}
\author{F. Godel}
\affiliation{Unit\'e Mixte de Physique, CNRS, Thales, Universit\'e Paris-Sud, Universit\'e Paris-Saclay, 91767, Palaiseau, France}
\author{P. R. Kidambi} 
\affiliation{Department of Chemical and Biomolecular Engineering, Vanderbilt University, 2400 Highland Avenue, Nashville, Tennessee 37212, USA}
\affiliation{Department of Engineering, University of Cambridge, Cambridge CB3 0FA, UK}
\author{S. Hofmann}
\affiliation{Department of Engineering, University of Cambridge, Cambridge CB3 0FA, UK} 
\author{X. P. Zhang}
\affiliation{Centro de F\'isica de Materiales (CFM-MPC) Centro Mixto CSIC-UPV/EHU,
20018 Donostia-San Sebasti\'an, Basque Country, Spain}
\affiliation{Donostia International Physics Center, 20018 Donostia-San Sebasti\'an, Spain}
\author{D. Bercioux}
\affiliation{Donostia International Physics Center, 20018 Donostia-San Sebasti\'an, Spain}
\affiliation{IKERBASQUE, Basque Foundation for Science, Maria Diaz de Haro 3, 48013 Bilbao, Spain}
\author{F. S. Bergeret}
\affiliation{Centro de F\'isica de Materiales (CFM-MPC) Centro Mixto CSIC-UPV/EHU,
20018 Donostia-San Sebasti\'an, Basque Country, Spain}
\affiliation{Donostia International Physics Center, 20018 Donostia-San Sebasti\'an, Spain}
\author{B. Dlubak}
\affiliation{Unit\'e Mixte de Physique, CNRS, Thales, Universit\'e Paris-Sud, Universit\'e Paris-Saclay, 91767, Palaiseau, France}
\author{P. Seneor}
\affiliation{Unit\'e Mixte de Physique, CNRS, Thales, Universit\'e Paris-Sud, Universit\'e Paris-Saclay, 91767, Palaiseau, France}
\author{Javier E. Villegas}
\email{javier.villegas@cnrs-thales.fr}
\affiliation{Unit\'e Mixte de Physique, CNRS, Thales, Universit\'e Paris-Sud, Universit\'e Paris-Saclay, 91767, Palaiseau, France}

\date{\today}
\begin{abstract}
Recent experiments have shown that proximity with high-temperature superconductors induces unconventional superconducting correlations in graphene. Here we demonstrate that those correlations propagate hundreds of nanometer, allowing for the unique observation of $d$-wave Andreev pair interferences in YBa$_2$Cu$_3$O$_7$-graphene devices that behave as a Fabry-P\'erot cavity. The interferences show as a series of pronounced conductance oscillations analogous to those originally predicted by de Gennes--Saint-James for conventional metal-superconductor junctions. The present work is pivotal to the study of exotic directional effects expected for nodal superconductivity in Dirac materials.  
\end{abstract}
\maketitle

The superconducting proximity effect in graphene has attracted much interest since the pioneering experiments~\cite{Heersche_2007}. This roots down to the graphene's electronic structure, which strongly affects the underlying mechanisms: Andreev reflection and coherent propagation of electron-hole pairs~\cite{Klapwijk_2004}. A distinctive feature is the strong dependence of the proximity behavior on the graphene?s doping level, which dramatically changes the Andreev reflection~\cite{Beenakker_2006,Efetov_2015} as compared to metals. Other unique features include gate~\cite{Allen_2015} or magnetic-field~\cite{Ben_Shalom_2015,Amet_2016} driven transitions from bulk to edge transport. Studies on graphene have also paved the way for understanding the proximity effect in other Dirac materials, such as topological insulators~\cite{Koren_2013,Finck_2014,Finck_2016,Bocquillon_2016}.

Experiments have mostly focused on conventional low critical temperature ($T_\text{C}$) superconductors with $s$-wave pairing. Despite early theoretical studies showing that $d$-wave (high-$T_\text{C}$) superconductors should lead to novel directional effects~\cite{Linder_2007,Linder_2008} and exotic pairing~\cite{Linder_2009,Awoga_2018}, evidence for unconventional superconductivity in graphene has been found only recently~\cite{Di_Bernardo_2017,Perconte_2017}. Scanning  electron tunneling microscopy (STM) of  graphene on Pr$_{2-x}$Ce$_{x}$CuO$_{4}$ (PCCO) revealed a superconducting gap~\cite{Di_Bernardo_2017} and spectral features suggesting ($p$-wave) superconductivity induced in graphene. Experiments on YBa$_{2}$Cu$_{3}$O$_{7}$ (YBCO) and chemical-vapor-deposited (CVD) graphene devices by some of us~\cite{Perconte_2017} showed transparent superconductor-graphene interfaces and clear evidence of Andreev reflection.  Interestingly, we also found that the Andreev electron-holes pair transmission can be modulated by a back-gate voltage, through a mechanism analogous to the Klein tunneling~\cite{Young_2009}. However, neither the STM experiments nor ours on solid-state devices probed the length scale over which the unconventional correlations penetrate into graphene.

%
%
\begin{figure}[!t]
		\centering
		\includegraphics[width=\columnwidth]{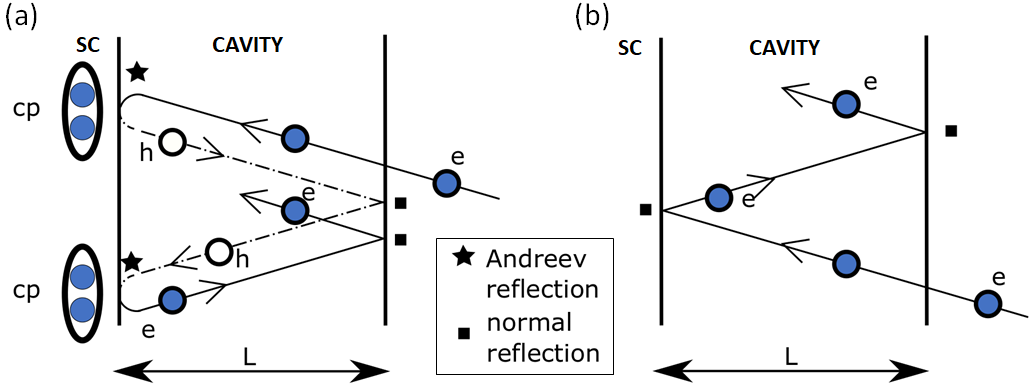}
		\caption{\label{figure1} Interferences in a proximitized cavity (a) An electron Andreev-reflected at the SC/cavity interface propagates as a hole back to the other end, where it is normal-reflected towards the SC/cavity interface to undergo again Andreev-reflection (AR). The lower AR is the time reversal process of the upper one. (b) The electron can also be normal-reflected at the SC/cavity, travel back to the other cavity interface to be again normal-reflected. This process results in Fabry-P\'erot resonances.}
\end{figure}
%
%
Here we demonstrate the long-range propagation of unconventional superconductivity into graphene via the observation of $d$-wave Andreev-pair interferences. These manifests themselves as conductance oscillations in devices that allow confining Andreev pairs within a graphene ``cavity''  whose length  $L$  is up to a few hundreds of nanometers. Predicted for proximitized graphene homojunctions~\cite{Linder_2007,Linder_2008}, the oscillations are analogous to the
de Gennes--Saint-James~\cite{de_Gennes_1963}  and McMillan--Rowel resonances~\cite{Rowell_1966} in the electronic density of states of ultrathin normal-metals backed by superconductors. Figure~\ref{figure1}(a) displays a cartoon of the underlying mechanism. Electrons injected in the cavity are Andreev-reflected as holes at the interface with the superconductor, retrace their path to the opposite cavity's end where they are normal-reflected, and travel bain ck towards the superconductor, where they are again Andreev-reflected (now as electrons). This results in destructive/constructive interferences dictated by the energy-dependent phase accumulated along the loop. This phenomenon shows up as conductance oscillations as a function of the bias voltage $V_\text{BIAS}$. Because Andreev pairs stemming from YBCO have $d$-wave symmetry, they decay over distances comparable to the mean free path  $l$~\cite{Tsuneto_1962,Balatsky_2006}, the observation of Andreev-pair interferences implies ballistic or quasi-ballistic transport. Consistently, the Andreev-pair interferences are accompanied by normal-electron resonances that result from normal-reflections at the two cavity's ends [sketch in Fig.~\ref{figure1}(b)] and commensurability between $L$ and the electrons' wavelength~\cite{Liang_2001,Miao_2007,Young_2009,Campos_2012,Allen_2017}. These different resonances result in oscillations as a function of both the gate  $V_{G}$ and bias  $V_\text{BIAS}$, and can be distinguished from Andreev-pairs related ones by their distinct periodicity and voltage regime in which they dominate. To our knowledge, the concurrent observation of both resonances had been restricted to junctions between $s$-wave superconductors and topological insulators~\cite{Finck_2014,Finck_2016}, in which Andreev-pair interferences manifest at very low-temperatures as a function of the gate voltage  $V_{G}$ . Here they can be observed at higher temperatures and show in a rich series of oscillations as a function of  $V_\text{BIAS}$  because of the high-$T_\text{C}$ and large gap of YBCO (tens of meV) which, contrary to low-temperature superconductors~\cite{Giazotto_2001}, may enclose many orders of interference.  

%
%
\begin{figure}[h]
	\centering
		\includegraphics[width=\columnwidth]{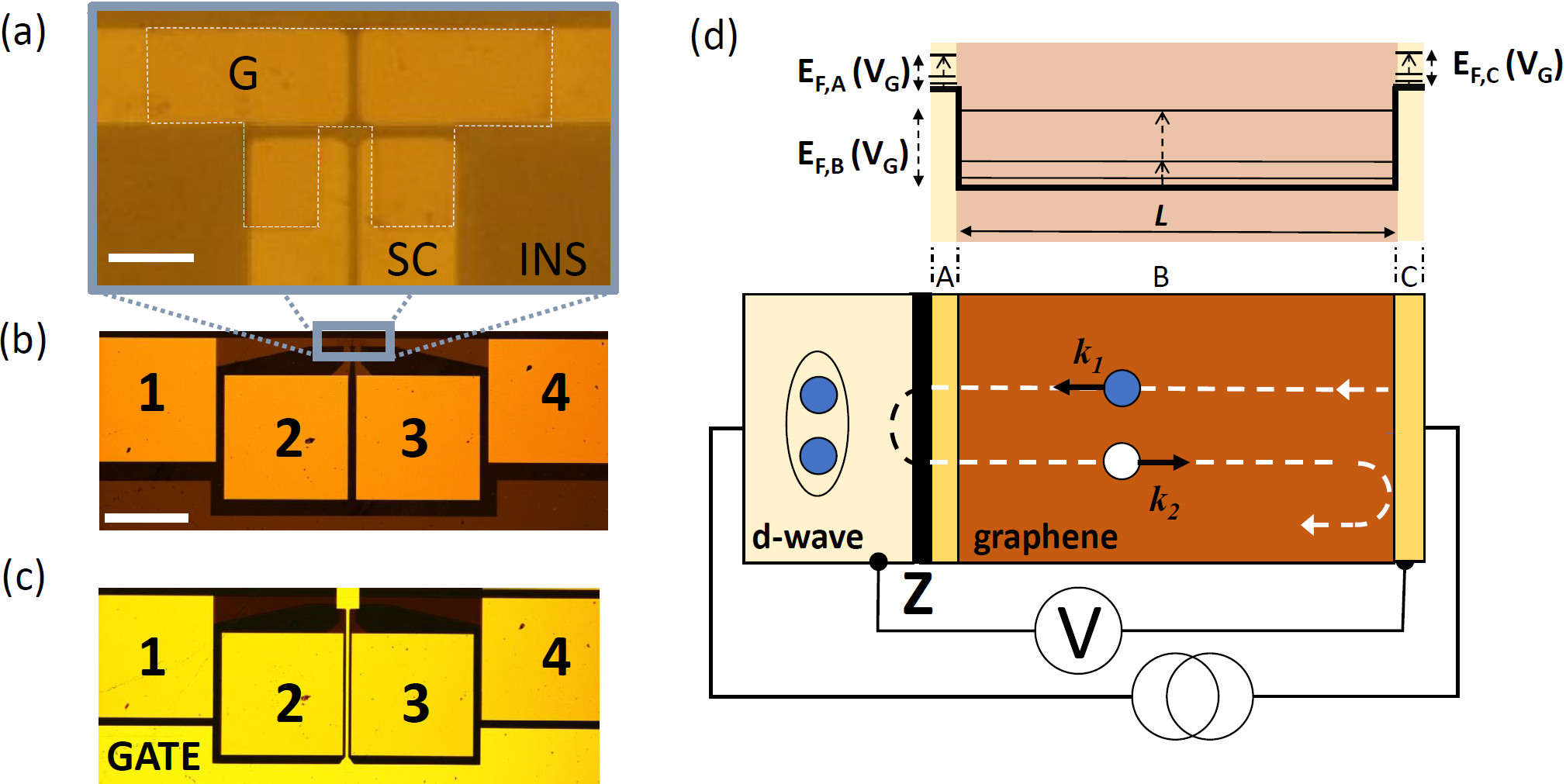}
		\caption{\label{figure2}Micrograph of a graphene/YBCO superconducting device. The scale bar is 10 $ \mu$m. (a) shows a zoom on the junction: Superconducting electrodes (SC) are connected by graphene bridges (G) on top of insulating YBCO (INS). The scale bar is 100~$\mu\text{m}$. (c) Micrograph of a device with the top gate. (d) Scheme of the graphene homojunction model. The doping depends on whether graphene lies on insulating or superconducting YBCO, which results in three graphene regions A B, C with different Fermi energy. This creates the cavity where interferences occur. The lower scheme shows the equivalent electrical circuit. The YBCO-Au-graphene interface, characterized by a barrier strength $Z$, is measured in series with the cavity.}	
\end{figure}
%
%
The planar devices [Fig.~\ref{figure2}(a)] consist of four superconducting (SC) c-axis oriented YBCO$_\text{50nm}$/Au$_\text{4nm}$ electrodes disposed within an insulating (INS) YBCO matrix and separated by a gap of length 100~nm$<L<$ 800~nm. As-grown YBCO films showed  $T_\text{C}\sim 90$~K while $T_\text{C}\sim 70$~K for lithographed devices, indicating moderate deoxygenation during fabrication. The ultrathin Au interlayer is deposited \emph{in situ} to preserve the superconducting properties and  improve the interfacial transparency~\cite{Perconte_2017}. The Au interlayer is crucial:  none of the effects reported here could be observed in its absence. Notice that the Au layer thickness it is one order of magnitude thinner than the mean free path in Au, and consequently $d$-wave correlations expectedly propagate undisturbed across it~\cite{Tsuneto_1962,Balatsky_2006}. A single-layer CVD grown~\cite{Kidambi_2012} graphene bridge (G) connects the four electrodes, each of which terminates in a gold pad [1-4 in Fig.~\ref{figure2}(b)] used for wire-bonding. The transparency of the electrical contact between each SC electrode and the graphene varied largely, even within a single device. Here we report on effects that arise only at high transparency contacts, which is often the case for just one out of four contacts per device. The structure is covered by a 45 nm thick AlO$_{x}$ layer (gate dielectric)~\cite{Dlubak_2012}.  As reported earlier~\cite{Mzali_2016}, with this encapsulation the carrier mobility in the used CVD graphene reaches up to  $\mu  \sim  7000~\text{cm}^{2}~\text{V}^{-1}\text{s}^{-1}$. The top-gate is made of Au entirely [Fig.~\ref{figure2}(c)] and the gate capacitance as measured over insulating YBCO is  $C_{G} \sim  5\times 10^{11}  e~\text{cm}^{-2}~\text{V}^{-1}$ (with $e$ the electron charge). Further details are reported elsewhere~\cite{Perconte_2017}.  

%
%
\begin{figure*}[!th]
	\centering
		\includegraphics[width=\textwidth]{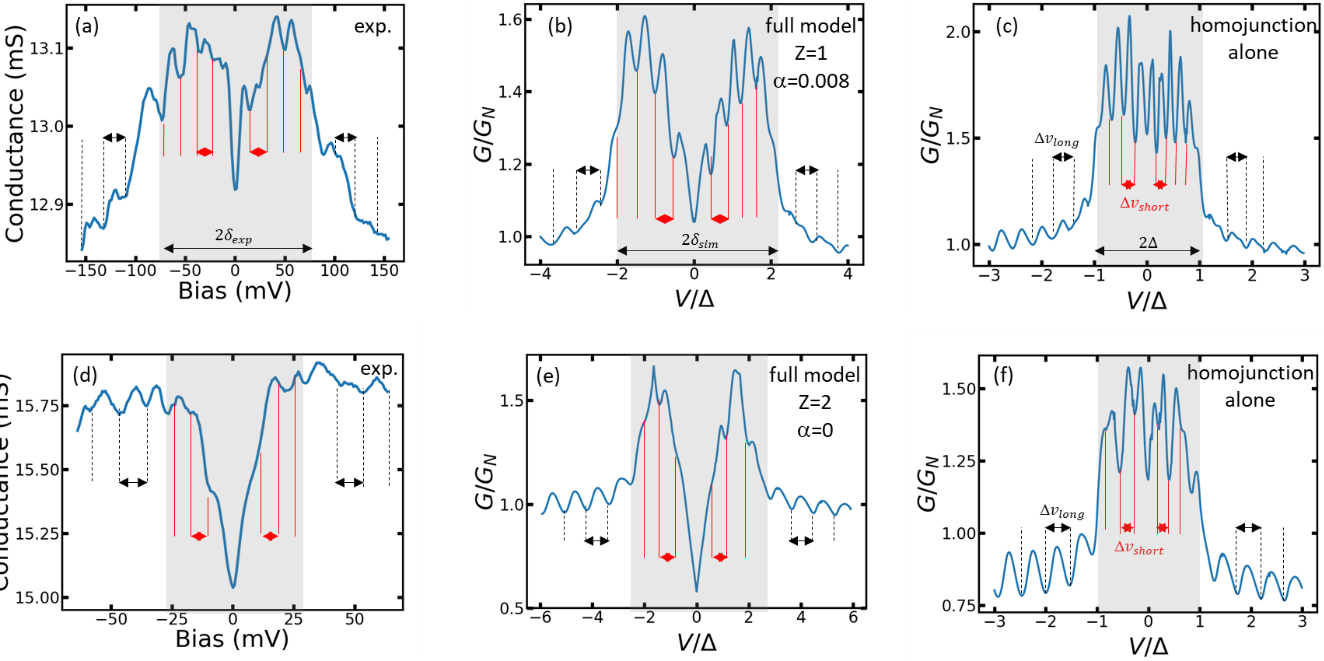}
		\caption{\label{figure3} Conductance versus $V_\text{BIAS}$ for two different devices, B3U (a) and B4U (b) measured at $3.2$~K. Vertical lines point out the series of oscillations, and the horizontal double-headed arrows indicate the periods  \( V_\text{long} \)   (black) and  $V_\text{short}$  (red). The shaded area indicates the width  $2 \delta$ of the superconducting-gap related feature. (b) and (e) show corresponding full-model calculations of the conductance with the $Z$ and indicated in the legend (c) and (f) shows simulations for a proximitized graphene cavity alone, without finite-$Z$ junction in series.}
\end{figure*}
%
%
We performed three-probe measurements, with the current $I$ injected \textit{e.g.} from 1 to contact 4 and the  $V_\text{BIAS}$ measured between 1 and 3. Thus we probe the YBCO-Au-graphene interface in series with the graphene ``cavity''  formed between the SC electrodes, as sketched in in Fig.~\ref{figure2}(d). The wiring as well as Au/YBCO contact resistances are Ohmic and negligible~\cite{Perconte_2017}. The differential conductance  $G(V_\text{BIAS}) =dI/dV_\text{BIAS}$  is measured using a Keithley 6221 current source coupled to a Keithley 2182.

\begin{figure}[!t]
	\centering
		\includegraphics[width=\columnwidth]{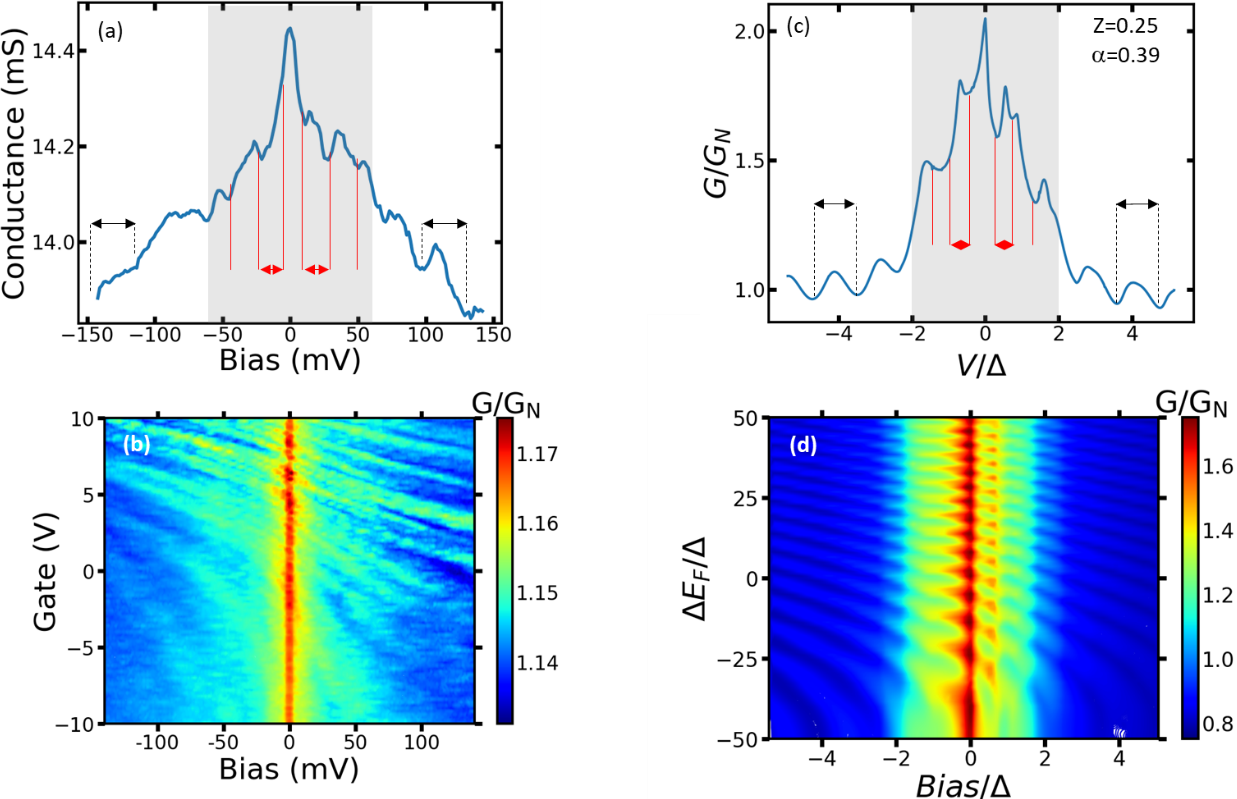}
		\caption{\label{figure4} (a) Conductance versus $V_\text{BIAS}$ for device A5U, measured at $T=4$~K. (b) Same device conductance (color scale) as function of bias voltage (horizontal axis) and gate voltage (vertical axis).  (c) Full-model calculation of the conductance with $Z$ and $\alpha$ indicated in the legend. (d) corresponding full-model calculation of the conductance as a function of bias and the Fermi energy variation in the cavity.}
\end{figure}

Examples of low-temperature ($3.2$~K) differential conductance measurements are shown in Fig.~\ref{figure3}(a),~\ref{figure3}(d) and Fig.~\ref{figure4}(a). The differential conductance  $G( V_\text{BIAS})$  shows a low-bias feature that stands out from the high-bias conductance level. That is either a conductance decrease [Fig.~\ref{figure3}(d)] or an enhancement [Fig.~\ref{figure4}(a)] within a typical bias range  $|V_\text{BIAS}|< \delta _{\exp} \sim~20-70$~mV highlighted by the grey shade. Those features are reminiscent of the conductance decrease/increase observed for  $e |V_\text{BIAS}|<\Delta$  in superconducting/normal-metal junctions depending on the interface transparency~\cite{Blonder_1982}. However, here they extend over a bias range  $\delta _{\exp}$ that clearly exceeds the superconducting energy-gap expected for moderately-underdoped to optimally-doped YBCO, which is reported in the range  $\Delta_\text{YBCO}\sim~15- 30$~meV~\cite{Maggio_Aprile_1995,Wei_1998,Dagan_2000,Rouco_2020}. 
Furthermore, details such as the sharp zero-bias peak observed for some devices (see \textit{e.g.} B4D and E3D in the Supplemental Material~\cite{SM}) are characteristic of junctions involving $d$-wave superconductors, and appear when the $d$-wave nodes form a certain angle with respect to the junction interface~\cite{Wei_1998,Kashiwaya_1995,Sharoni_2001}.

In addition to the central features, an oscillation pattern appears in Fig.~\ref{figure3}(a) and~\ref{figure3}(d), this extends over a bias range well above $\delta _{\exp}$. Indeed, two types of oscillations exist: the ones with longer period  $V_\text{long}$, predominant for  $\vert V_\text{BIAS}~ \vert > \delta _{\exp}$, and the other with shorter  $V_\text{short}$  which are more prominent at  $\vert V_\text{BIAS}~ \vert < \delta _{\exp}$.

Figure~\ref{figure4} show an example of gating effects. Figure~\ref{figure4}(a) displays  $G(V_\text{BIAS})$  without gate voltage ($V_{G}=0)$, and shows the general features discussed above. Figure~\ref{figure4}(b) is a color plot of a series of $G(V_\text{BIAS})$  at constant temperature and varying  $V_{G}$. The central (red) feature corresponds to the zero-bias conductance peak. The conductance is periodically modulated by $V_{G}$: a pattern of oblique lines (light-blue/green) indicates that  $V_{G}$  gradually ``shifts"  the oscillations observed as a function of  $V_\text{BIAS}$. The oblique lines' slope gradually varies with  $V_{G}$, which results in pronounced curvature over the plot periphery (low  $V_{G}$  or high  $|V_\text{BIAS}|$).

We discuss now a model that explains the main observations, namely (i) the increase or decrease of the zero-bias conductance; (ii) the origin and period of conductance oscillations. The model is based on the Blonder-Tinkham-Klapwijk (BTK) formalism~\cite{Blonder_1982} extended to junctions between $d$-wave superconductors and normal metals~\cite{Kashiwaya_1996} and to proximitized graphene homojunctions~\cite{Linder_2007,Linder_2008}. A scheme is shown in Fig.~\ref{figure2}(d). First, we consider the YBCO-Au-graphene interface. Because the Au thickness ($ \sim 5$~nm) is well-below the mean free path $l_\text{Au} \sim~40$~nm~\cite{Scheer_2001} and the low-$T$ coherence length  $\xi_\text{Au}=v_{F}l/(6\pi KT)\sim30$~nm~\cite{Sharoni_2004}, we characterize that interface [black in Fig.~\ref{figure2}(d)] via a single Blonder-Tinkham-Klapwijk barrier-strength parameter $Z$. If $Z=0$, the transmission between YBCO and graphene is only mediated by the Andreev reflection. By increasing $Z$, Andreev reflection turns less dominant, normal reflection is enhanced, and transmission is dominated by tunneling.  We model the graphene ``channel''  where Andreev pairs and normal electrons propagate, considering three different regions A, B, C. A and C correspond to graphene lying on superconducting YBCO/Au [yellow in Fig.~\ref{figure2}(d)], but we consider that only contact A has a relatively low $Z$.  The region B of length $L$ [dark in Fig.~\ref{figure2}(d)] lies on insulating YBCO. Here the Fermi energy  $E_{F,B}$   is expectedly different from  $E_{F,A}$,  $E_{F,C}$  as the graphene's doping depends on the substrate's electronic properties~\cite{Hwang_2012}. Within this model, the structure is a graphene homojunction with a gate-tunable Fermi energy step, as sketched in the upper Fig.~\ref{figure2}(d), and behaves as a resonant cavity for electrons and Andreev pairs.

%
%
%
%
The conductance is calculated numerically by considering the YBCO-Au-graphene interface~\cite{Wei_1998,Kashiwaya_1996}  and the proximitized graphene homojunction~\cite{Linder_2007,Linder_2008} in-series (see Supplemental Material for details~\cite{SM}). The input parameters are  $Z$, the reduced Fermi energies  $\epsilon_{F,i} \equiv E_{F,i}/ \Delta$  (with  $i  =\text{A, B, C}$ and  $\Delta$  the energy-gap induced in A by proximity effect), the reduced cavity's length  $\Lambda  \equiv L/ \lambda _{F,A}$  (with the Fermi wavelength  $\lambda _{F,A}$, chosen as a reference following the original model~\cite{Linder_2007,Linder_2008}), and  $\alpha$  the effective angle between $d$-wave anti-nodes and the homojunction interfaces. The key parameters are $\Lambda$  and  $\epsilon _{F,A}$ : they determine  $V_\text{long}$  and  $V_\text{short}$.  The parameters $\alpha$  and $Z$ determine the conductance background shape, and particularly whether there is a conductance decrease or enhancement around zero bias. We stress that the sharp zero-bias peak observed in some devices cannot be reproduced by the simulation unless we consider  $\alpha\neq 0$  (see Supplemental Material~\cite{SM} Sec.~7). This is a consequence of the $d$-wave symmetry of the correlations propagating across the different interfaces. The parameters $\epsilon_{F,B} $ and  $\epsilon _{F,C}$  only change the oscillations' phase and amplitude, but do not affect their period~\cite{Linder_2008}. To choose the simulation parameters that yield the best agreement with the experimental curves we proceed as detailed in the Supplemental Material in Sec.~2-4~\cite{SM}. Simulations examples are shown in Fig.~\ref{figure3}(b),~\ref{figure3}(e) and Fig.~\ref{figure4}(c) for the experiments in Fig.~\ref{figure3}(a),~\ref{figure3}(d) and Fig.~\ref{figure4}(a), respectively. The simulations closely reproduce the main experimental features [further examples and the simulations' parameters shown in the Supplemental Material~\cite{SM} in Secs.~5 and~6].

In order to illustrate the different contributions to the conductance, Figs.~\ref{figure3}(c) and~\ref{figure3}(f) display the conductance of a proximitized graphene homojunction~\cite{Linder_2008} alone, \textit{i.e.} without a finite-$Z$ junction in series. The same   $\epsilon _{F,i}$, $\Lambda$  and  $\alpha$  as for the full-model calculations are used. Here the conductance shows a background dependence on $V_\text{BIAS}$  very different from the experiments, which evidences that it is strongly influenced by the YBCO-Au-graphene interface. In particular, the zero-bias ``dip"  only emerges if a finite-$Z$ junction is considered.  However, Figs.~\ref{figure3}(c) and~\ref{figure3}(f) do show the short and long period oscillations observed experimentally. This demonstrates that they originate within the graphene homojunction. Notice that the ratio between the short and long periods\ is different from that in the experiments (and full model simulations): in Fig.~\ref{figure3}(c) and~\ref{figure3}(f), $V_\text{short}$  is clearly shorter (relative to  $V_\text{long}$) than in Fig.~\ref{figure3}(a)-\ref{figure3}(b) and~\ref{figure3}(d)-\ref{figure3}(e). This is because in experiments (and full-model)  $V_\text{BIAS}$  is divided between the graphene homojunction and the YBCO-Au-graphene interface, and the strongly non-linear conductance leads to a voltage distribution that varies depending on the  $V_\text{BIAS}$. In particular, for  $eV_\text{BIAS}< \Delta$  the conductance across the YBCO-Au-graphene interface decreases (this is more pronounced for higher $Z$ values) while the homojunction conductance increases [see Figs.~\ref{figure3}(c),~\ref{figure3}(f)]. Consequently,  $V_\text{short}$  (observed for   $eV_\text{BIAS}< \Delta$) appears ``stretched''  relative to the   $V_\text{long}$  (observed for $eV_\text{BIAS}> \Delta$).

\begin{figure}[!t]
		\includegraphics[width=\columnwidth]{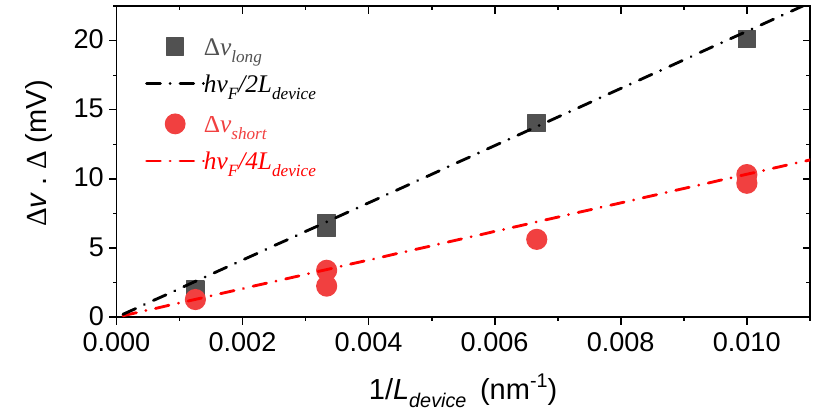}
		\caption{\label{figure5} Oscillations period as a function of the cavity length  $L_\text{device}$. The black and red dashed lines respectively correspond to the period theoretically expected for electron (Fabry-P\'erot) and Andreev-pair (De Gennes-Saint James) interferences.}
\end{figure}

The oscillations' physical meaning can be understood from Fig.~\ref{figure5}.  This displays the short (red circles) and long (black squares) oscillation periods as a function of the inverse of the cavity length  $L_\text{device}^{-1}$ defined upon devices fabrication. Specifically, we plot the period  $V=v   \Delta$ , where $\Delta$ is estimated as discussed in the Supplemental Material~\cite{SM} and  $v$  is the period obtained from our model once the contribution of the YBCO-Au-graphene interface has been removed --- as in Fig.~\ref{figure3}(c) and~\ref{figure3}(e) --- to avoid the aforementioned  $V_\text{BIAS}$  division artifacts.

The long-period oscillations follow $V_{\text{th, long}}=hv_{F}/2L_\text{device}$. This is as expected from the interference between electrons travelling back and forth from one cavity side to the other after normal reflections [Fig.~\ref{figure1}(b)]. This period results~\cite{Liang_2001,Miao_2007,Young_2009,Campos_2012,Allen_2015}  
 from the interference condition  $2L_\text{device}k=2n \pi$  (with  $n$  an integer and  $k$  the electron wavevector) and the graphene's linear dispersion, which yields  $V=\hbar v_{F}k$~\cite{Castro_Neto_2009}. These interferences imply normal-electron coherence over $ \sim L_\text{device}$. Their amplitude is greater for $100~\text{nm} <  L_\text{device} < 300$~nm but drastically diminishes beyond that. This suggests that the mean free path is above 100~nm but clearly below 1~$\mu$m, in agreement with a rough estimate based on the carrier mobility~\cite{Hwang_2008} which yields  $l\lesssim 100- 260$~nm . Notice that the scaling of the period with $1/  L_\text{device}$  rules out Tomasch resonances~\cite{Tomasch_1966,Nesher_1999,Visani_2012} within YBCO, which should scale with its thickness and should therefore show the same period for all devices.

%
%
%
%
The short period oscillations, clearly visible for  $V< \Delta$, follow  $V_{th,short}=hv_{F}/4L_\text{device}$. This period is expected from the interference of Andreev pairs~\cite{de_Gennes_1963,Rowell_1966} --- scheme in Fig.~\ref{figure1}(a). In this case the resonance condition reads  $2L_\text{device} \vert k_{1}-k_{2} \vert =2n \pi$, where  $k_{1}$  and  $k_{2}$  are the electron and reflected hole wavevectors, whose difference is established by the applied voltage  $\vert k_{1}-k_{2} \vert =2V/\hbar v_{F}$~\cite{Rowell_1966}. These resonances imply that the superconducting coherence is preserved over $\sim L_\text{device}$. 

Our model also explains the gating effects. Figure~\ref{figure4}(d) displays a simulated conductance as a function of bias and Fermi energy, which reproduces the main experimental features --- particularly the pattern of oblique lines that characterize the conductance modulation by by  $V_{G}$  [Fig.\ref{figure4}(b)]. The correspondence between the \textit{y}-axis in simulations and experiments implies that the Fermi energy (and vector  $ k_{F}$) is varied about  proportionally to  $V_{G}$, in agreement with earlier experiments~\cite{Perconte_2017}. Varying  $k_{F}$  produces a periodic modulation of the conductance, via the resonance condition  $2L_\text{device}k_{F}=2n\pi$, yielding the oblique lines. Note that the model also reproduces the  slope gradual change, and the pronounced curvature over the plot's periphery, which is obtained by including in the model  $\epsilon _{F,B} (V_{G})\neq\epsilon_{F,A}(V_{G})=\epsilon_{F,C}(V_{G})$  to account for a partial pining of graphene Fermi energy on conducting YBCO.

In summary, we have realized ballistic planar devices in which graphene is proximitized by a $d$-wave cuprate superconductor. We find that the $d$-wave correlations propagate into graphene over hundreds of nanometers at $\sim 4$~K. The confinement of Andreev pairs and electrons in graphene homojunctions of submicrometric lateral dimensions produces quantum interferences that show as conductance oscillations. Those involving Andreev pairs, more prominent below the superconducting gap, are analogous to the De Gennes-Saint James oscillations. Although difficult to observe in graphene/low-temperature superconductor devices, here they are very clear due to the large YBCO gap (tens of mV), which fits many orders of interference within. The unusual d-wave Andreev-pair interferences are accompanied by normal-electron ones. The  simultaneous observation of both  is rare, and stems from the fact that the coherence length of d-wave correlations is limited by the mean free path~\cite{Balatsky_2006}. Thus, the necessary condition for both types of resonances is satisfied simultaneously here. Extensions of this study to graphene with lower intrinsic doping and higher carrier mobility should allow exploring specular Andreev reflection  regimes~\cite{Beenakker_2006} rand Josephson effects. The latter would also require enhancing the transparency of the graphene/YBCO interface, possibly by replacing the Au interlayer, in order to routinely obtain low-$Z$ junctions. Further, perspectives include novel directional effects~\cite{Linder_2009,Awoga_2018} linked to the $d$-wave character of the induced correlations, which could be exploited in more sophisticated devices to realize topological states~\cite{Lucignano_2013}. This work should also further encourage studies in which graphene is replaced by other 2D materials with  intrinsically interesting proximity behavior~\cite{Kuzmanovski_2016,Linder_2017}, eventually combined with exfoliated $d$-wave superconductors~\cite{Li_2017}.

\begin{acknowledgements}
Work at Unit\'e Mixte de Physique CNRS/Thales supported by the ERC Grant No.  647100 ``SUSPINTRONICS'', French ANR Grants No. ANR-15-CE24-0008-01 ``SUPERTRONICS''  and No. ANR-17-CE30-0018-04 ``OPTOFLUXONICS'', Labex NanoSaclay No. ANR-10-LABX-0035, European COST action 16218 ``Nanocohybri'' , and EU Work Programme Graphene Flagship (Grants No. 696656 and 785219). This work was supported by the French RENATECH network (French national nanofabrication platform). X.P.Z., D.B., and F.S.B. acknowledge support from Ministerio de Ciencia e Innovacion  (MICINN) through Project No. FIS2017-82804-P.  F. S. B. acknowledges Grupos Consolidados UPV/EHU del Gobierno Vasco (Grant No. IT1249- 19). We thank Professor. J. Santamar\'ia for discussions.
\end{acknowledgements}

\bibliography{bibliography}
\end{document}


\title{Supplemental Material:\\Long-Range Propagation and Interference of $d$-wave Superconducting Pairs in Graphene}
\author{D. Perconte}
\affiliation{Unit\'e Mixte de Physique, CNRS, Thales, Universit\'e Paris-Sud, Universit\'e Paris-Saclay, 91767, Palaiseau, France}
\affiliation{Laboratorio de Bajas Temperaturas y Altos Campos Magn\'eticos, Departamento de F\'isica de la Materia Condensada, Instituto Nicol\'as Cabrera and Condensed Matter Physics Center (IFIMAC), Universidad Aut\'onoma de Madrid, E-28049 Madrid, Spain}
\author{K. Seurre}
\affiliation{Unit\'e Mixte de Physique, CNRS, Thales, Universit\'e Paris-Sud, Universit\'e Paris-Saclay, 91767, Palaiseau, France}
\author{V. Humbert} 
\affiliation{Unit\'e Mixte de Physique, CNRS, Thales, Universit\'e Paris-Sud, Universit\'e Paris-Saclay, 91767, Palaiseau, France}
\author{C. Ulysse}
\affiliation{Centre for Nanoscience and Nanotechnology, CNRS, Universit\'e Paris-Sud/Universit\'e Paris-Saclay, Boulevard Thomas Gobert, Palaiseau, France}
\author{A. Sander} 
\affiliation{Unit\'e Mixte de Physique, CNRS, Thales, Universit\'e Paris-Sud, Universit\'e Paris-Saclay, 91767, Palaiseau, France}
\author{J. Trastoy} 
\affiliation{Unit\'e Mixte de Physique, CNRS, Thales, Universit\'e Paris-Sud, Universit\'e Paris-Saclay, 91767, Palaiseau, France}
\author{V. Zatko} 
\affiliation{Unit\'e Mixte de Physique, CNRS, Thales, Universit\'e Paris-Sud, Universit\'e Paris-Saclay, 91767, Palaiseau, France}
\author{F. Godel}
\affiliation{Unit\'e Mixte de Physique, CNRS, Thales, Universit\'e Paris-Sud, Universit\'e Paris-Saclay, 91767, Palaiseau, France}
\author{P. R. Kidambi} 
\affiliation{Department of Chemical and Biomolecular Engineering, Vanderbilt University, 2400 Highland Avenue, Nashville, Tennessee 37212, USA}
\affiliation{Department of Engineering, University of Cambridge, Cambridge CB3 0FA, UK}
\author{S. Hofmann}
\affiliation{Department of Engineering, University of Cambridge, Cambridge CB3 0FA, UK} 
\author{X. P. Zhang}
\affiliation{Centro de F\'isica de Materiales (CFM-MPC) Centro Mixto CSIC-UPV/EHU,
20018 Donostia-San Sebasti\'an, Basque Country, Spain}
\affiliation{Donostia International Physics Center, 20018 Donostia-San Sebasti\'an, Spain}
\author{D. Bercioux}
\affiliation{Donostia International Physics Center, 20018 Donostia-San Sebasti\'an, Spain}
\affiliation{IKERBASQUE, Basque Foundation for Science, Maria Diaz de Haro 3, 48013 Bilbao, Spain}
\author{F. S. Bergeret}
\affiliation{Centro de F\'isica de Materiales (CFM-MPC) Centro Mixto CSIC-UPV/EHU,
20018 Donostia-San Sebasti\'an, Basque Country, Spain}
\affiliation{Donostia International Physics Center, 20018 Donostia-San Sebasti\'an, Spain}
\author{B. Dlubak}
\affiliation{Unit\'e Mixte de Physique, CNRS, Thales, Universit\'e Paris-Sud, Universit\'e Paris-Saclay, 91767, Palaiseau, France}
\author{P. Seneor}
\affiliation{Unit\'e Mixte de Physique, CNRS, Thales, Universit\'e Paris-Sud, Universit\'e Paris-Saclay, 91767, Palaiseau, France}
\author{Javier E. Villegas}
\email{javier.villegas@cnrs-thales.fr}
\affiliation{Unit\'e Mixte de Physique, CNRS, Thales, Universit\'e Paris-Sud, Universit\'e Paris-Saclay, 91767, Palaiseau, France}
\date{\today}
\maketitle
\tableofcontents
\section{Theoretical model}
We reproduce the experimental device conductance via numerical simulations based on combination of the models of Refs.~\cite{Linder_2007} and~\cite{Wei_1998}.  The model of Ref.~\cite{Linder_2007} applies to a graphene homojunction proximitized by a $d$-wave superconductor. The junction is divided into three different sections\ A, B, C connected in series.  Each of them has a different Fermi energy and together form an energy quantum well which is tuneable upon application of a gate voltage (in the case studied here, we assume that the different gate capacitance on insulating and metallic YBCO lead to different gating effects in A, C and B). In order to allow for Andreev reflection at the A/B interface, the model~\cite{Linder_2007} assumes that  the electronic density of states in region A presents a superconducting energy-gap. The homostructure conductance is obtained by matching the electronic wave functions in the A/B and C/B interfaces. Given the transmission   $ T= \left( 1-|r|^{2} \right)$  and reflection  $R=|r_{A}|^{2}$  coefficient, the full expressions for $r$  and  $r_{A}$  are cumbersome and can be found in the appendix of Ref.~\cite{Linder_2008}. The conductance across the A/B/C structure is given by:
%
%
\begin{equation}\label{eq_s1}
 G_{1} \left( V_{1} \right) = \int _{-\frac{ \pi }{2}}^{\frac{ \pi }{2}}d \theta  \left( 1-|r|^{2} \right) \cos  \left(  \theta  \right) +|r_{A}|^{2}\cos\left(  \theta _{A} \right). 
\end{equation}
%
%
With  $\theta _{A}=\arcsin  \left[ \left( eV_{1}+E_{f} \right) / \left( eV_{1}-E_{f} \right) \sin\left(  \theta  \right)  \right]$  and  $\theta$  the electron angle of incidence with respect to the interfaces. Both  $r$ and  $r_{A}$  and thus the conductance  $G_{1}(V_1)$ depend on the following parameters: the reduced channel length  $\Lambda  \equiv L/ \lambda _{F,A}$  (with  $\lambda _{F,A}$ the Fermi wavelength in A),  the reduced graphene Fermi energy in the different regions ($\epsilon _{F,i} \equiv E_{F,i}/ \Delta$,   with $i$=A, B, C) and  $\Delta$  the amplitude of the superconducting energy-gap, and the angle between the $d$-wave nodes of the superconducting order parameter and the interface  $\alpha$. The model makes a full wave matching calculation such that effects coming from different angles of incidence after several reflections are taken into account. As discussed in the manuscript, this first block of the model explains the conductance oscillations as a function of gate and bias voltage. It is important to notice that the above result applies to electrons that propagate ballistically within the B-region. 

The second block of the model takes into account the finite transmission of the YBCO/Au/graphene interface that is model as a $d$-wave/metal interface as studied in Refs.~\cite{Wei_1998} and~\cite{Perconte_2017}. Specifically, the conductance is given by:
%
%
\begin{equation}\label{eq_s2}
G_{2} \left( V_{2} \right) = \int_{-\frac{\pi}{2}}^{\frac{\pi}{2}} d\theta  \frac{16 \left( 1+ \vert  \Gamma _{+} \vert ^{2} \right) \cos  (\theta) ^4+4Z^2 \left( 1- \vert  \Gamma _{+} \Gamma _{-} \vert ^{2}\right)  \cos  (\theta)^2}{ \vert 4\cos  (  \theta  ) ^{2}+Z^{2} \left( 1- \Gamma _{+} \Gamma _{-} \right) \vert^{2}},
\end{equation}
%
%
where  $\Gamma _\pm= \left( eV_{2}/ \vert  \Delta\left( \theta _\pm \right)  \vert  \right) - \sqrt{\left( eV_{2}/ \vert  \Delta\left(  \theta _\pm \right)  \vert  \right) -1}$  and  $\theta _{+}= \theta$, $\theta _{-}= \pi - \theta$ . While  $\Delta(\theta) = \Delta \cos\left[ 2 \left(\theta - \alpha\right)\right]$  and   $\alpha$  is the angle between the superconducting order parameter and the interface. The conductance  \( G_{2} \left( V_{2} \right)  \)  thus depends on the parameters   $\alpha$  and on the BTK barrier strength  $Z$. As discussed in the main text this block of the model leads to a proper description of the background conductance vs. the voltage bias.

Using  $I= \int _{}^{}G \left( V \right) dV$  we obtain  $G_{1} \left( I \right)$ and   $G_{2} \left( I \right)$ and  from these, we calculate the conductance when the two building blocks are connected in-series:
%
%
\begin{equation}\label{eq_s3}
 G \left( I \right) =\frac{1}{ G_{1}(I)^{-1} +G_{2}(I)^{-1}} 
\end{equation}
%
%
Finally, using  $V= \int\frac{1}{G(I)}dI$, we obtain  $G(V)$.

\section{Simulation procedure} 

In order to reproduce the experimental curves via simulations, we typically follow the procedure illustrated below. Blue curves are experimental data and orange curves correspond to the simulations. The examples in the following is is based on device A3D.

We first find $Z$ and  $\alpha$  that allow mimicking the conductance background. These parameters respectively characterize the YBCO/Au-graphene interface and the $d$-wave nature of the superconducting correlations.

\begin{figure}[!h]
	\centering
	\begin{minipage}[c]{0.8\textwidth}
		\centering
		\includegraphics[width=0.39\textwidth]{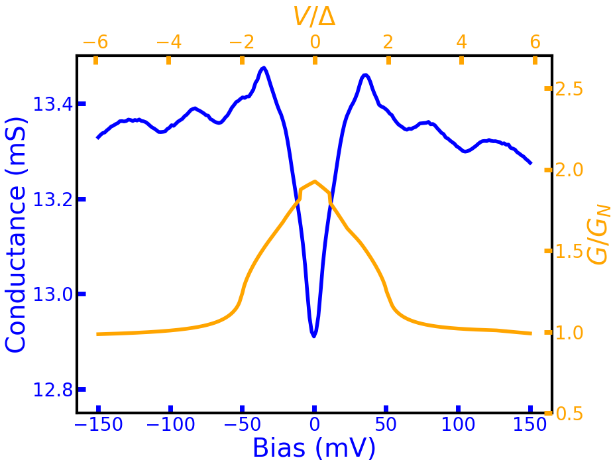}
		\includegraphics[width=0.39\textwidth]{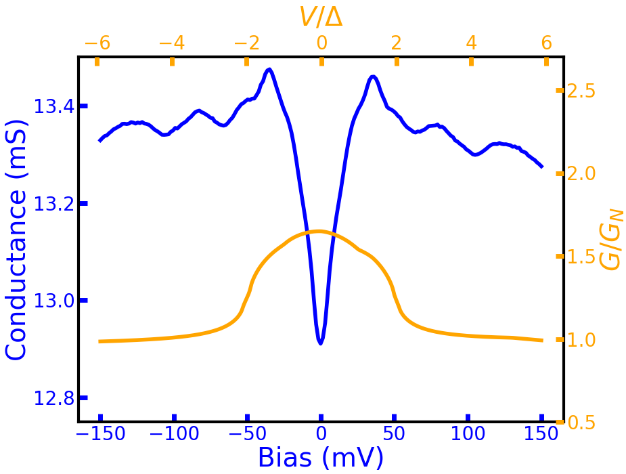}
	\end{minipage}
	\begin{minipage}[c]{0.8\textwidth}
		\centering
		\includegraphics[width=0.39\textwidth]{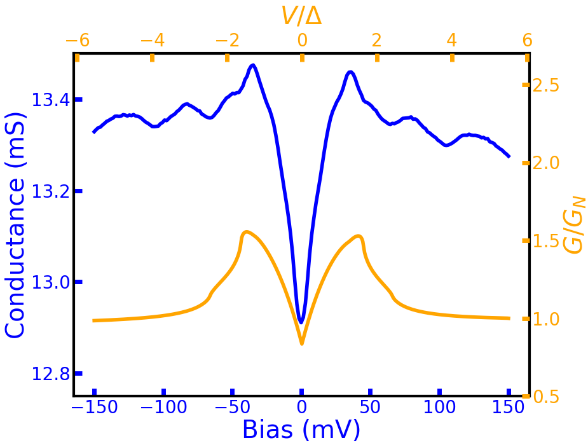}
		\includegraphics[width=0.39\textwidth]{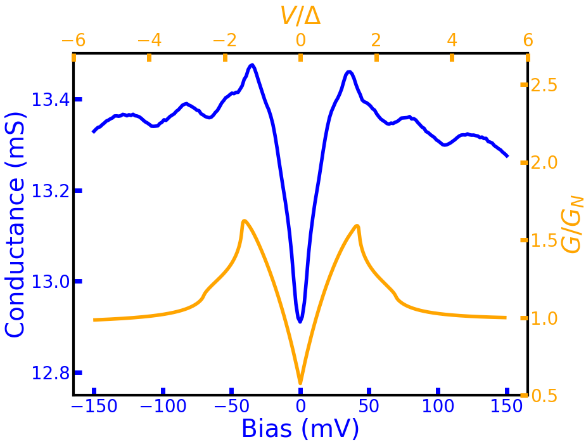}
	\end{minipage}
\caption{\label{Fig_one} Differential conductance experimental (blue) and theoretical (orange) as a function of the bias for $\Lambda\to 0$, $\alpha=0$ and various values of the $Z$ parameter: $Z=0$ (top left), $Z=0.5$ (top right), $Z=1$ (bottom left) and $Z=2$ (bottom right).}
\end{figure}

Let us first assume that the normalized cavity length $\Lambda\to 0$  (so that there are no visible oscillations), fix $\alpha =0$  and vary $Z$, the results are presented in the various panels of Fig.~\ref{Fig_one}. We see that one needs relatively large $Z=2$ to better reproduce the V-shaped ``dip"  around zero bias. 

\begin{figure}[!h]
	\centering

	\begin{minipage}[c]{0.8\textwidth}
		\centering
		\includegraphics[width=0.39\textwidth]{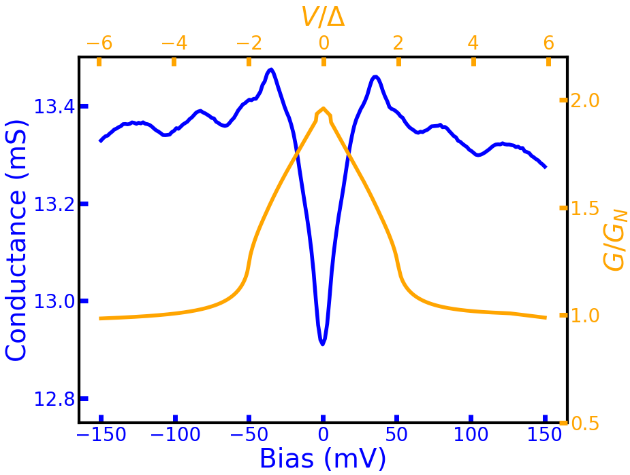}
		\includegraphics[width=0.39\textwidth]{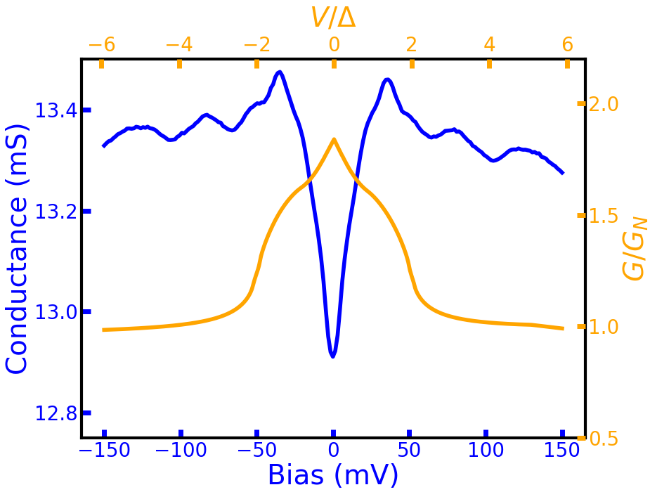}
	\end{minipage}
	\begin{minipage}[c]{0.8\textwidth}
		\centering
		\includegraphics[width=0.39\textwidth]{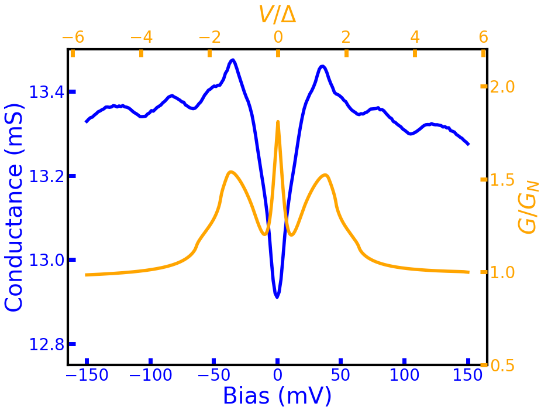}
		\includegraphics[width=0.39\textwidth]{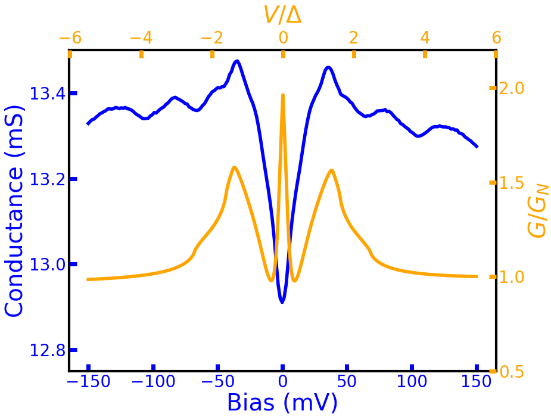}
	\end{minipage}
\caption{\label{Fig_two} Differential conductance experimental (blue) and theoretical (orange) as a function of the bias for $\alpha=0.23$ and various values of the $Z$ parameter: $Z=0$ (top left), $Z=0.5$ (top right), $Z=1$ (bottom left) and $Z=2$ (bottom right).}
\end{figure}

\begin{figure}[!h]
	\centering
	\begin{minipage}[c]{0.8\textwidth}
		\centering
		\includegraphics[width=0.39\textwidth]{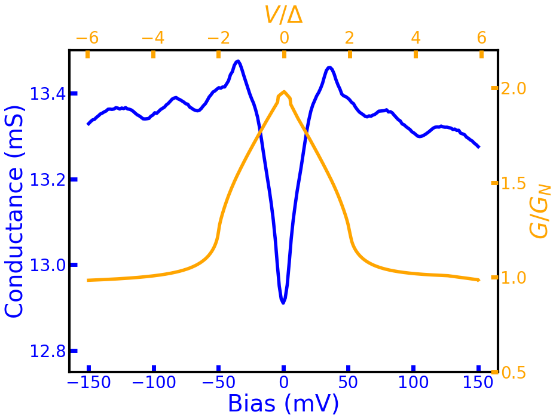}
		\includegraphics[width=0.39\textwidth]{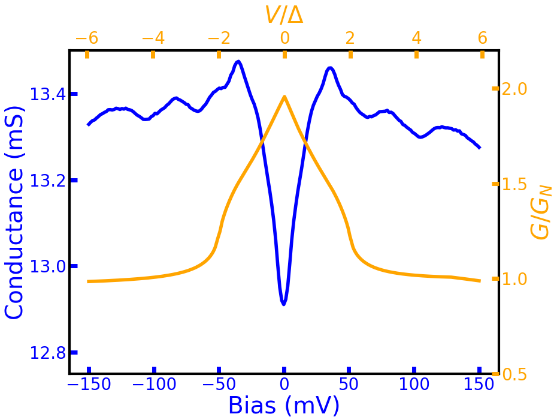}
	\end{minipage}
	\begin{minipage}[c]{0.8\textwidth}
		\centering
		\includegraphics[width=0.39\textwidth]{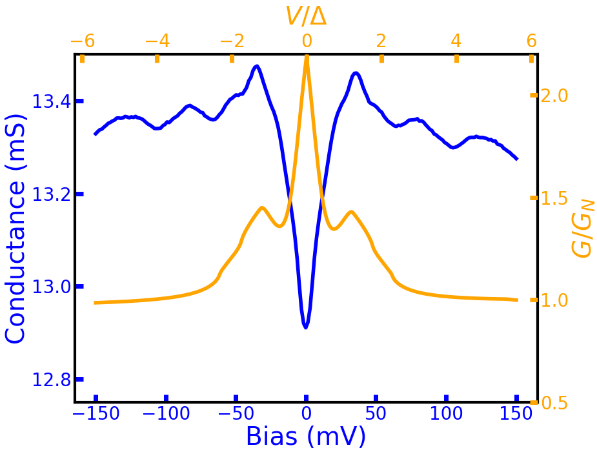}
		\includegraphics[width=0.39\textwidth]{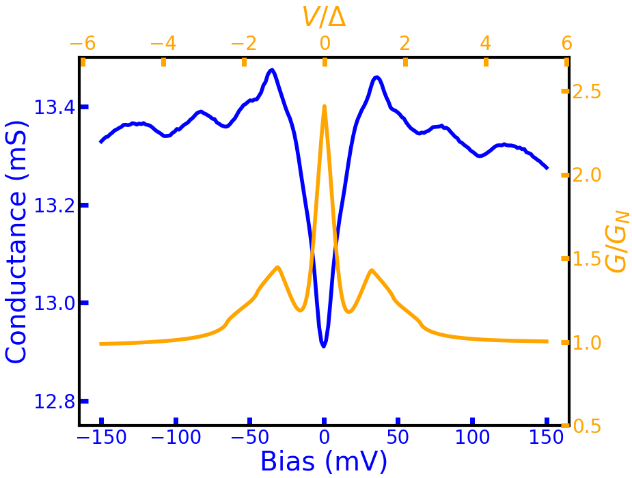}
	\end{minipage}
\caption{\label{Fig_three} Differential conductance experimental (blue) and theoretical (orange) as a function of the bias for $\alpha=0.39$ and various values of the $Z$ parameter: $Z=0$ (top left), $Z=0.5$ (top right), $Z=1$ (bottom left) and $Z=2$ (bottom right).}
\end{figure}

Now we explore the effect of varying  $\alpha$ in Figs.~\ref{Fig_two} and~\ref{Fig_three}. With these examples, one quickly realizes that it is not possible to use  $\alpha  \neq 0$  in the present case, regardless of the choice of $Z$, because  $\alpha\neq 0$ yields to a sharp zero/bias conductance peak that is absent for this particular device.
%
%
%
%
%
%
%
%
%
%
%
%

\noindent Considering the above, we would fix  $Z=2$  and  $\alpha =0$  for the A3D device. 

Now we look for  $\Lambda$  and $\epsilon_{F,A}$ that yield oscillations of the same reduced period   $\Delta V/ \delta$  as in the experiments. We show in Figs.~\ref{Fig_four} and~\ref{Fig_five} some examples that illustrate how these parameters influence the oscillations period. First, we show examples with fixed  $\epsilon_{F,A}$  and varying  $\Lambda$  (the red vertical lines mark minima in the experimental curve, allowing for a reference of experimental high-bias period):
%
%
\begin{figure}[!h]
	\centering
	\begin{minipage}[c]{0.8\textwidth}
		\centering
		\includegraphics[width=0.39\textwidth]{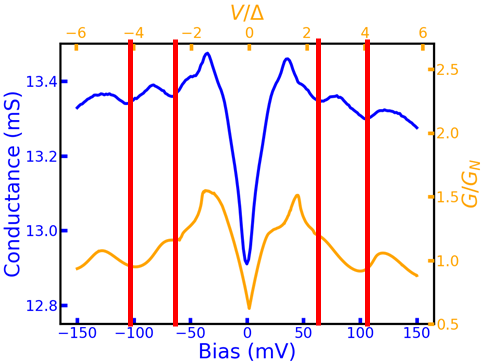}
		\includegraphics[width=0.39\textwidth]{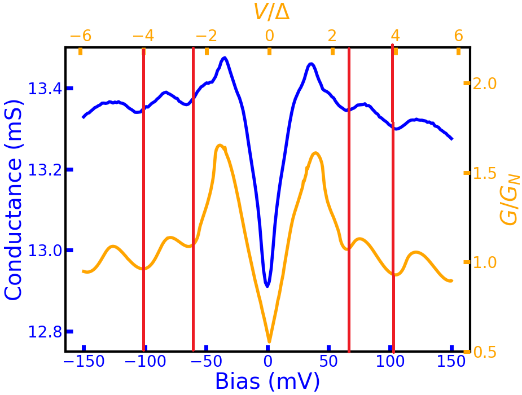}
	\end{minipage}
	\begin{minipage}[c]{0.8\textwidth}
		\centering
		\includegraphics[width=0.39\textwidth]{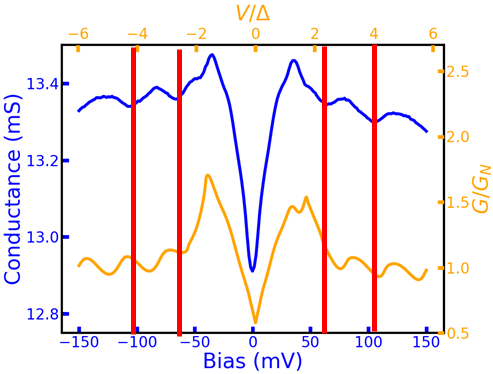}
		\includegraphics[width=0.39\textwidth]{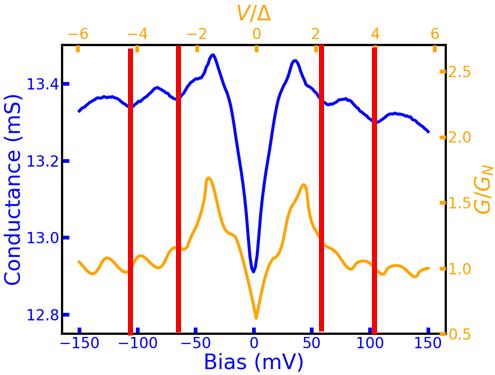}
	\end{minipage}
\caption{\label{Fig_four} Differential conductance experimental (blue) and theoretical (orange) as a function of the bias for $\epsilon_{F,A}=20$ and various values of the $\Lambda$ parameter: $\Lambda=8$ (top left), $\Lambda=10$ (top right), $\Lambda=13$ (bottom left) and $\Lambda=16$ (bottom right).}
\end{figure}
%
%
%
%
\begin{figure}[!h]
	\centering
	\begin{minipage}[c]{0.8\textwidth}
		\centering
		\includegraphics[width=0.39\textwidth]{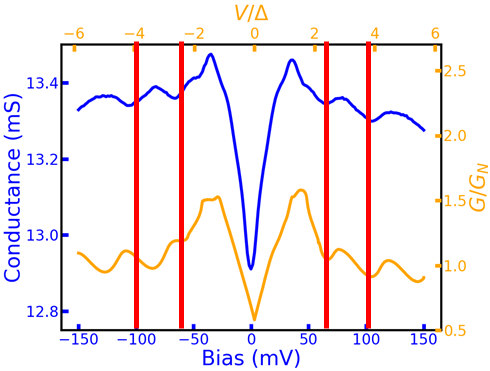}
		\includegraphics[width=0.39\textwidth]{image14.png}
	\end{minipage}
	\begin{minipage}[c]{0.8\textwidth}
		\centering
		\includegraphics[width=0.39\textwidth]{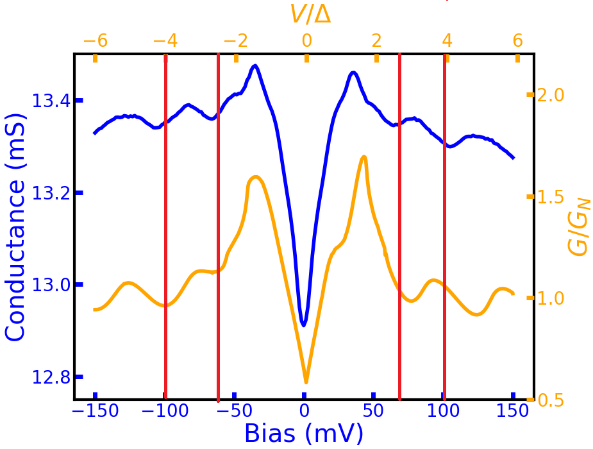}
		\includegraphics[width=0.39\textwidth]{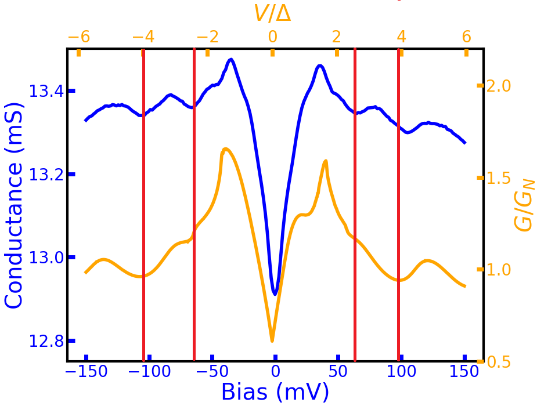}
	\end{minipage}
\caption{\label{Fig_five} Differential conductance experimental (blue) and theoretical (orange) as a function of the bias for $\Lambda=10$ and various values of the $\epsilon_{F,A}$ parameter: $\epsilon_{F,A}=18$ (top left), $\epsilon_{F,A}=20$ (top right), $\epsilon_{F,A}=22$ (bottom left) and $\epsilon_{F,A}=25$ (bottom right).}
\end{figure}
%
%
We see that increasing the cavity length  $\Lambda$  diminishes the oscillations period, as one would naturally expect. As shown in the Figs.~\ref{Fig_four} and~\ref{Fig_five}, if one fixes  $\Lambda$  and varies  $\epsilon_{F,A}$  the period is also modified, which is also expected. Indeed, from the definition of the parameters and the graphene's linear dispersion relation~\cite{Linder_2008}, the oscillations period is the same for any constant  $\Lambda/\epsilon _{F,A}$. Note however that the choice of  $\epsilon_{F,A}$  is not entirely free because this parameter modifies may introduce an asymmetry in the curve. Globally the best agreement is found for the combination  $\epsilon _{F,A}=20$,  $\Lambda =10$. 

\begin{figure}[!h]
	\centering
	\begin{minipage}[c]{0.8\textwidth}
		\centering
		\includegraphics[width=0.39\textwidth]{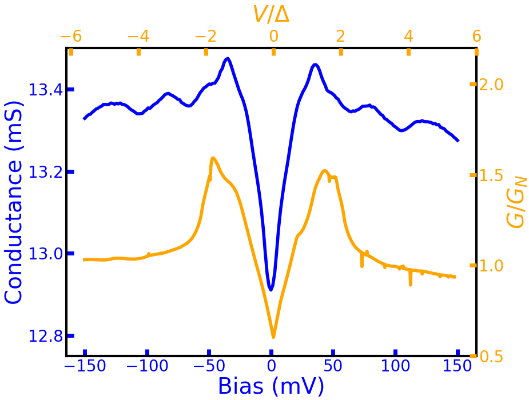}
		\includegraphics[width=0.39\textwidth]{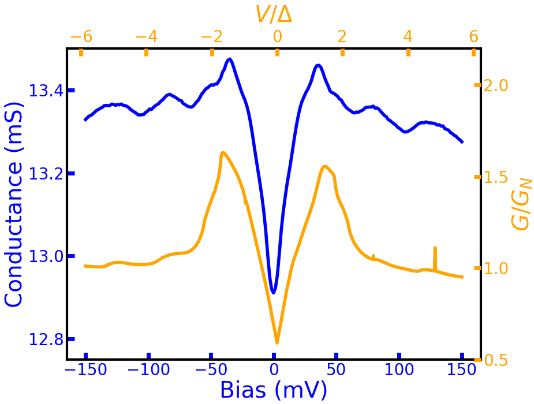}
	\end{minipage}
	\begin{minipage}[c]{0.8\textwidth}
		\centering
		\includegraphics[width=0.39\textwidth]{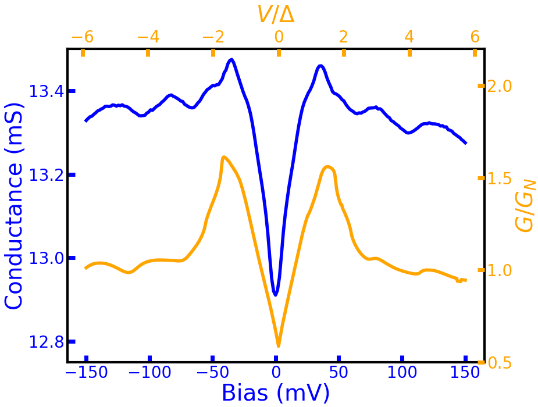}
		\includegraphics[width=0.39\textwidth]{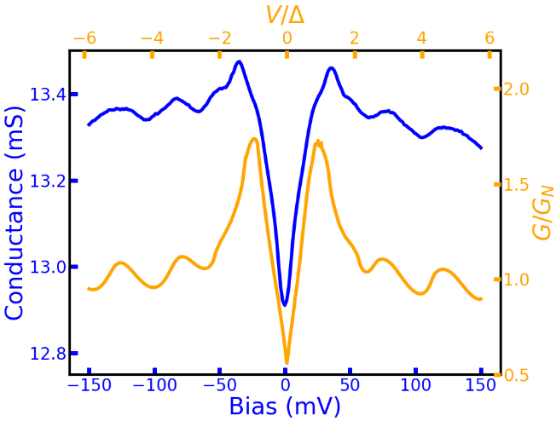}
	\end{minipage}
\caption{\label{Fig_six} Differential conductance experimental (blue) and theoretical (orange) as a function of the bias at the optimal parameter point and various values of the $\epsilon_{F,B}$ parameter: $\epsilon_{F,B}=50$ (top left), $\epsilon_{F,B}=100$ (top right), $\epsilon_{F,B}=150$ (bottom left) and $\epsilon_{F,B}=400$ (bottom right).}
\end{figure}

\begin{figure}[!h]
	\centering
	\begin{minipage}[c]{0.8\textwidth}
		\centering
		\includegraphics[width=0.39\textwidth]{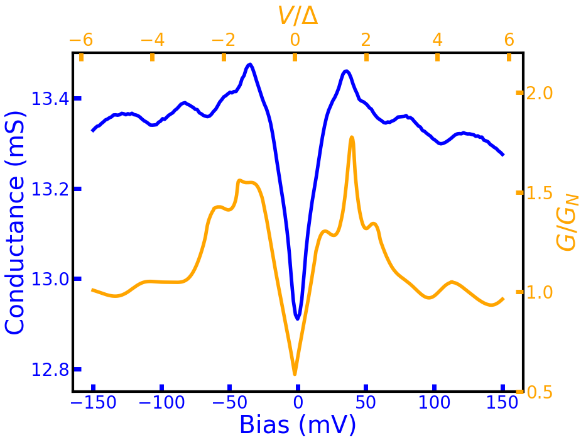}
		\includegraphics[width=0.39\textwidth]{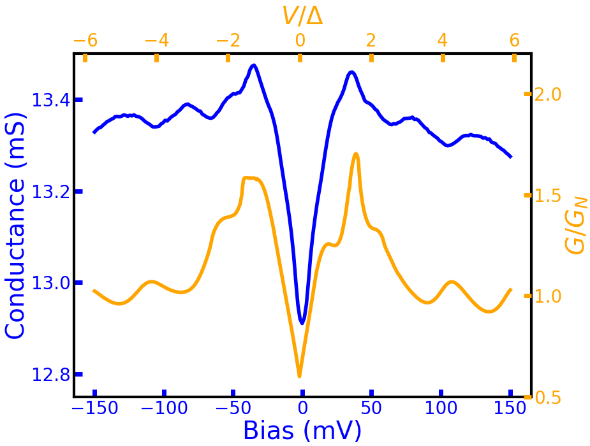}
	\end{minipage}
	\begin{minipage}[c]{0.8\textwidth}
		\centering
		\includegraphics[width=0.39\textwidth]{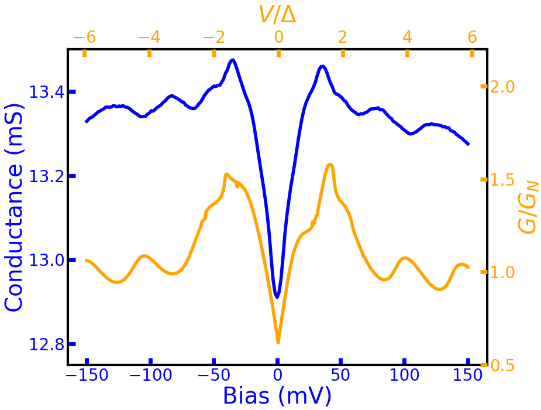}
		\includegraphics[width=0.39\textwidth]{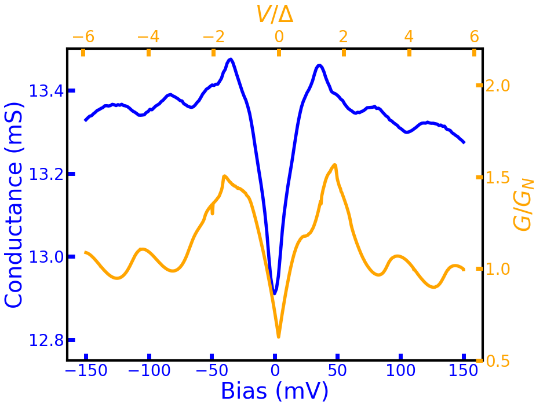}
	\end{minipage}
\caption{\label{Fig_seven} Differential conductance experimental (blue) and theoretical (orange) as a function of the bias at the optimal parameter point and various values of the $\epsilon_{F,C}$ parameter: $\epsilon_{F,C}=8$ (top left), $\epsilon_{F,C}=12$ (top right), $\epsilon_{F,C}=16$ (bottom left) and $\epsilon_{F,C}=19$ (bottom right).}
\end{figure}

Finally, $\epsilon_{F,B}$  and  $\epsilon_{F,C}$  are adjusted in order to modulate the oscillations amplitude. As shown earlier for this model~\cite{Linder_2008}, the two latter parameters have no impact on the oscillations period, only on their amplitude and phase. This is because the Fermi velocity mismatch modifies the transmission/reflection across the interface. This is illustrated in Figs.~\ref{Fig_six} and~\ref{Fig_seven} where all the other parameters are kept constant at the optimum values deduced above.
%
%
%
%
%
%
%
%

\clearpage
\section{Variability of the fitting parameters }

In order to further illustrate the effect and variability of the key simulation parameters, we include below examples that show their impact on the simulations.

\begin{figure}[h]
		\includegraphics[width=0.5\textwidth]{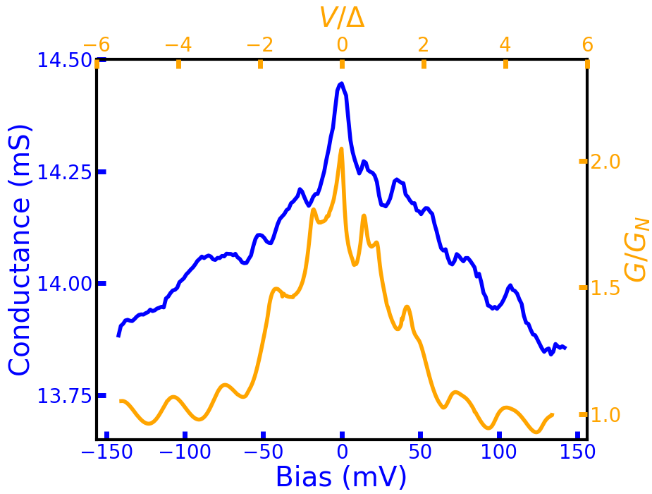}
		\caption{\label{Fig_eigth} Device \textbf{A5U}: Differential conductance experimental (blue) and theoretical (orange) as a function of the bias for $\epsilon_{F,A}=14$, $Z=0.25$, $\alpha=0.39$.}
\end{figure}

First examples: device, A5U. Best agreement simulation/experiment for  $\epsilon _{F,A}=14$ ,  $Z=0.25$,   $\alpha =0.39$, cf Fig.~\ref{Fig_eigth}.

%
%
%
%

%
%
\begin{figure}[!th]
	\centering
	\begin{minipage}[c]{0.8\textwidth}
		\centering
		\includegraphics[width=0.39\textwidth]{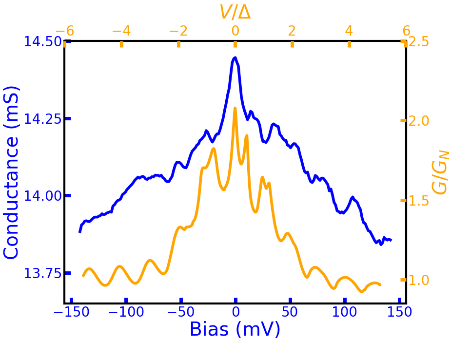}
		\includegraphics[width=0.39\textwidth]{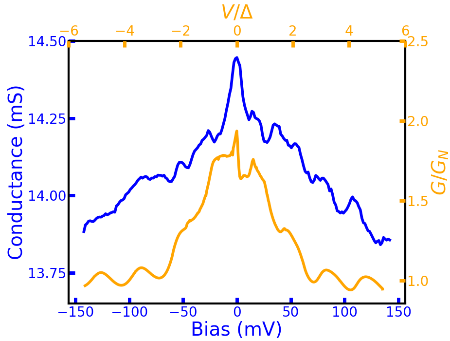}
	\end{minipage}
	\begin{minipage}[c]{0.8\textwidth}
		\centering
		\includegraphics[width=0.39\textwidth]{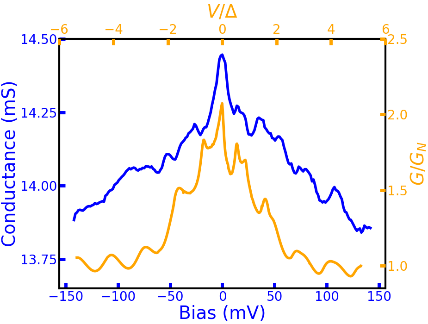}
		\includegraphics[width=0.39\textwidth]{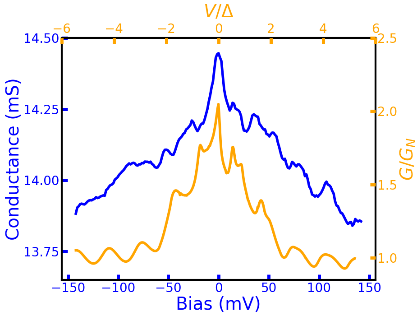}
	\end{minipage}
	\begin{minipage}[c]{0.8\textwidth}
		\centering
		\includegraphics[width=0.39\textwidth]{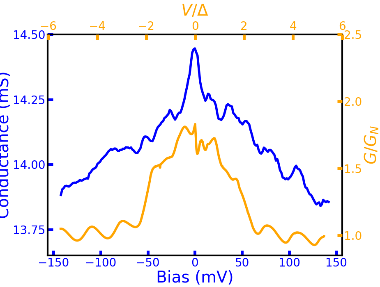}
		\includegraphics[width=0.39\textwidth]{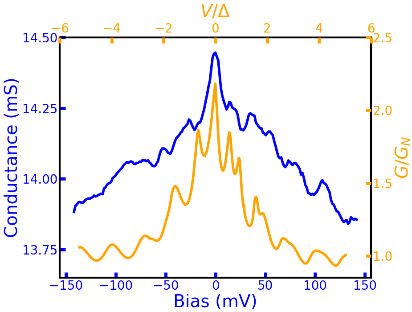}
	\end{minipage}
\caption{\label{Fig_nine} Differential conductance experimental (blue) and theoretical (orange) as a function of the bias for different sets of fitting parameters: $\epsilon_{F,A}=12$, $Z=0.25$, $\alpha=0.39$ (top left), $\epsilon_{F,A}=16$, $Z=0.25$, $\alpha=0.39$ (top right), $\epsilon_{F,A}=14$, $Z=0$, $\alpha=0.39$ (central left), $\epsilon_{F,A}=14$, $Z=0.5$, $\alpha=0.39$ (central right), $\epsilon_{F,A}=14$, Z=0.25, $\alpha=0.23$ (bottom left), $\epsilon_{F,A}=14$, $Z=0.25$, $\alpha=0.53$ (bottom right).}
\end{figure}
%
%

\begin{figure}[h]
		\includegraphics[width=0.5\textwidth]{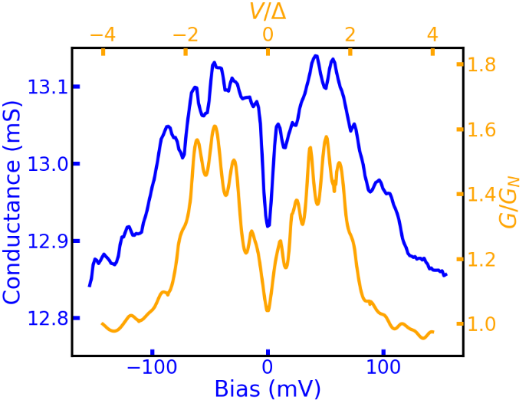}
		\caption{\label{Fig_ten} Device \textbf{B3U}: Differential conductance experimental (blue) and theoretical (orange) as a function of the bias for $\epsilon_{F,A}=20$, $Z=1$, $\alpha=0.008$.} 
\end{figure}

Second example: device B3U. Best agreement for,   $\epsilon _{F,A}=20$,  $Z=1$,  $\alpha =0.008$, cf. Fig.~\ref{Fig_ten}.

%
%
\begin{figure}[!h]
	\centering
	\begin{minipage}[c]{0.8\textwidth}
		\centering
		\includegraphics[width=0.35\textwidth]{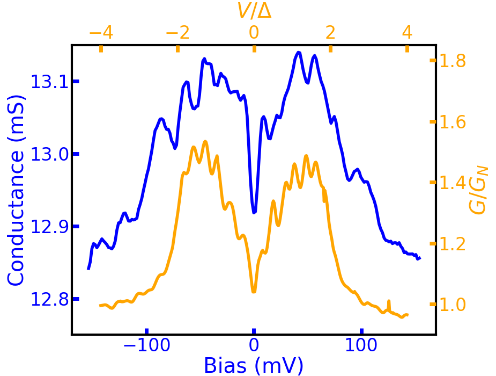}
		\includegraphics[width=0.35\textwidth]{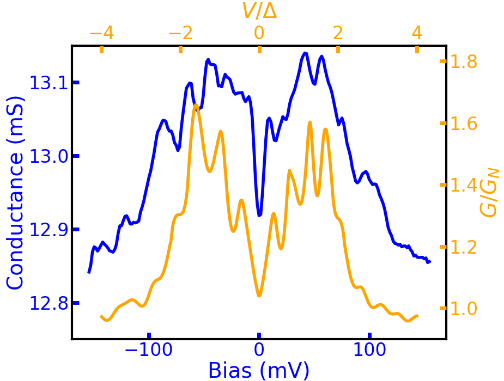}
	\end{minipage}
	\begin{minipage}[c]{0.8\textwidth}
		\centering
		\includegraphics[width=0.35\textwidth]{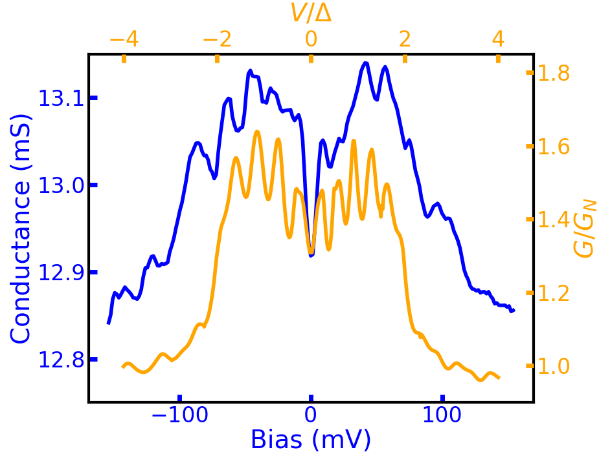}
		\includegraphics[width=0.35\textwidth]{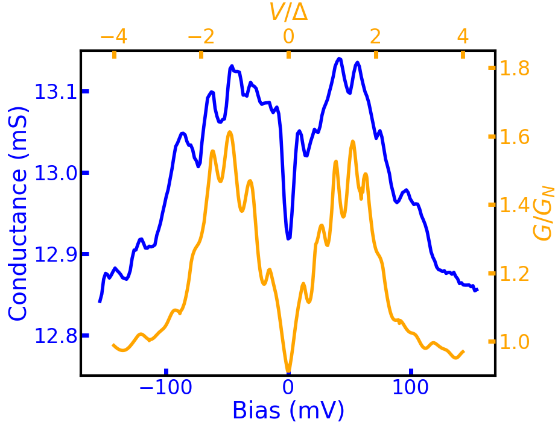}
	\end{minipage}
	\begin{minipage}[c]{0.8\textwidth}
		\centering
		\includegraphics[width=0.35\textwidth]{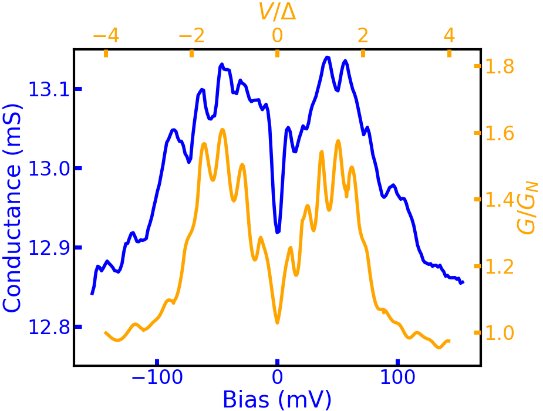}
		\includegraphics[width=0.35\textwidth]{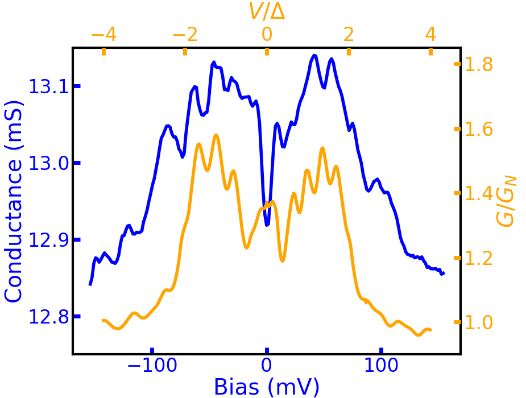}
	\end{minipage}
\caption{\label{Fig_eleven} Differential conductance experimental (blue) and theoretical (0.39) as a function of the bias for different sets of fitting parameters: $\epsilon_{F,A}=15$, $Z=1$, $\alpha=0.008$ (top left), $\epsilon_{F,A}=25$, $Z=1$, $\alpha=0.008$ (top right), $\epsilon_{F,A}=20$, $Z=0.7$, $\alpha=0.008$ (central left), $\epsilon_{F,A}=20$, $Z=1.25$, $\alpha=0.008$ (central right), $\epsilon_{F,A}=20$, $Z=1$, $\alpha=0$ (bottom left), $\epsilon_{F,A}=20$, $Z=1$, $\alpha=0.157$ (bottom right).}
\end{figure}
%
%
From the these examples, one can see that the parameters can be typically varied  $Z \pm 0.25,~  \alpha  \pm 0.1$  and   $\epsilon_{F,A} \pm 1$  before a significant change is observable in the simulations.

\clearpage
\section{Quantitative evaluation of the simulation parameters }

The simulations are based on dimensionless parameters whose values for each device are given in a table below. It is very important that their consistency can be quantitatively evaluated using fundamental relations and the actual device dimensions. As explained, above the two parameters that determine the oscillations period are  $\Lambda$  are $\epsilon_{F,A}$. Specifically, what defines the oscillations period is the ratio $\Lambda /\epsilon_{F,A}$. 

If we take the definition of the parameters  $\Lambda  \equiv L/ \lambda _{F,A}$  and  $\epsilon _{F,A} \equiv E_{F,A}/ \Delta$,  and we consider the dispersion relationship in graphene  $E_{F}=v_{F}k_{F}=hv_{F}/ \lambda _{F}$  (with  $k_{F}$  and  $v_{F}=10^{6}~\text{m}\cdot\text{s}^{-1}$ the Fermi vector and velocity, and  $h$ the Planck constant), it is straightforward to find the following expression for the superconducting gap  $\Delta$:  
%
%
\begin{equation}\label{eq_s4}
\Delta = \left( \frac{ \Lambda }{ \epsilon _{F,A}} \right) hv_{F}L^{-1}
\end{equation}
%
%

Equation~\eqref{eq_s4} allows evaluating the simulations parameters quantitatively. If we input in Eq.~\eqref{eq_s4} the actual graphene cavity length in the studied devices  $L_{device}$  (distance between YBCO electrodes defined lithographically) and the used  $\frac{\Lambda}{\epsilon_{F,A}}$, we obtain a value for superconducting gap, which is similar for most of the devices varies between $15$~meV and $20$~meV. This is well within the values expected for the same YBCO films when moderately underdoped~\cite{Rouco_2020} and materials in proximity with YBCO~\cite{Sharoni_2004}. 

That a strong proof of the consistency of the analysis, from the quantitative point of view.
\section{Table of simulation parameters}

%
%
\begin{table}[h]
 	\centering
	\begin{tabular}{P{0.47in}P{0.47in}P{0.47in}P{0.47in}P{0.47in}P{0.47in}P{0.47in}P{0.23in}P{1.01in}}
	\hline
		\multicolumn{1}{|P{0.47in}}{ \textbf{Device}} & 
		\multicolumn{1}{|P{0.47in}}{ $\bm{L_{device}}$ } & 
		\multicolumn{6}{|P{\dimexpr3.58in+10\tabcolsep\relax}}{ \textbf{Input parameters}} & 
		\multicolumn{1}{|P{1.01in}|}{ \bm{$\Delta$} \textbf{ } \par  \textbf{from Eq.~\ref{eq_s4}}} \\
		\hline
		\multicolumn{1}{|P{0.47in}}{} & 
		\multicolumn{1}{|P{0.47in}}{{\fontsize{10pt}{12.0pt}\selectfont \textit{nm}}} & 
		\multicolumn{1}{|P{0.47in}}{  \textbf{Z}} & 
		\multicolumn{1}{|P{0.47in}}{$\bm{\alpha}$ } & 
		\multicolumn{1}{|P{0.47in}}{$\bm{\Lambda}$} & 
		\multicolumn{1}{|P{0.47in}}{$\bm{\epsilon_{F,A}}$ } & 
		\multicolumn{1}{|P{0.47in}}{$\bm{\epsilon_{F,B}}$ } & 
		\multicolumn{1}{|P{0.23in}}{$\bm{\epsilon_{F,C}}$ } & 
		\multicolumn{1}{|P{1.01in}|}{{\fontsize{10pt}{12.0pt}\selectfont \textit{meV}}} \\
		\hline
		\multicolumn{1}{|P{0.47in}}{ {\fontsize{10pt}{12.0pt}\selectfont A3D}} & 
		\multicolumn{1}{|P{0.47in}}{ {\fontsize{10pt}{12.0pt}\selectfont 100}} & 
		\multicolumn{1}{|P{0.47in}}{ {\fontsize{10pt}{12.0pt}\selectfont 2}} & 
		\multicolumn{1}{|P{0.47in}}{ {\fontsize{10pt}{12.0pt}\selectfont 0}} & 
		\multicolumn{1}{|P{0.47in}}{ {\fontsize{10pt}{12.0pt}\selectfont 10}} & 
		\multicolumn{1}{|P{0.47in}}{ {\fontsize{10pt}{12.0pt}\selectfont 20}} & 
		\multicolumn{1}{|P{0.47in}}{ {\fontsize{10pt}{12.0pt}\selectfont 400}} & 
		\multicolumn{1}{|P{0.23in}}{ {\fontsize{10pt}{12.0pt}\selectfont 17.5}} & 
		\multicolumn{1}{|P{1.01in}|}{ {\fontsize{10pt}{12.0pt}\selectfont 20.7}} \\
		\hline
		\multicolumn{1}{|P{0.47in}}{ {\fontsize{10pt}{12.0pt}\selectfont A3U}} & 
		\multicolumn{1}{|P{0.47in}}{ {\fontsize{10pt}{12.0pt}\selectfont 100}} & 
		\multicolumn{1}{|P{0.47in}}{ {\fontsize{10pt}{12.0pt}\selectfont 2}} & 
		\multicolumn{1}{|P{0.47in}}{  {\fontsize{10pt}{12.0pt}\selectfont 0}} & 
		\multicolumn{1}{|P{0.47in}}{  {\fontsize{10pt}{12.0pt}\selectfont 38}} & 
		\multicolumn{1}{|P{0.47in}}{  {\fontsize{10pt}{12.0pt}\selectfont 40}} & 
		\multicolumn{1}{|P{0.47in}}{  {\fontsize{10pt}{12.0pt}\selectfont 42.5}} & 
		\multicolumn{1}{|P{0.23in}}{  {\fontsize{10pt}{12.0pt}\selectfont 15.5}} & 
		\multicolumn{1}{|P{1.01in}|}{  {\fontsize{10pt}{12.0pt}\selectfont 39.3}} \\
		\hline
		\multicolumn{1}{|P{0.47in}}{  {\fontsize{10pt}{12.0pt}\selectfont A5U}} & 
		\multicolumn{1}{|P{0.47in}}{  {\fontsize{10pt}{12.0pt}\selectfont 150}} & 
		\multicolumn{1}{|P{0.47in}}{  {\fontsize{10pt}{12.0pt}\selectfont 0.25}} & 
		\multicolumn{1}{|P{0.47in}}{  {\fontsize{10pt}{12.0pt}\selectfont 0.39}} & 
		\multicolumn{1}{|P{0.47in}}{  {\fontsize{10pt}{12.0pt}\selectfont 10}} & 
		\multicolumn{1}{|P{0.47in}}{  {\fontsize{10pt}{12.0pt}\selectfont 14}} & 
		\multicolumn{1}{|P{0.47in}}{  {\fontsize{10pt}{12.0pt}\selectfont 150}} & 
		\multicolumn{1}{|P{0.23in}}{  {\fontsize{10pt}{12.0pt}\selectfont 9.5}} & 
		\multicolumn{1}{|P{1.01in}|}{  {\fontsize{10pt}{12.0pt}\selectfont 19.7}} \\
		\hline
		\multicolumn{1}{|P{0.47in}}{  {\fontsize{10pt}{12.0pt}\selectfont B3U}} & 
		\multicolumn{1}{|P{0.47in}}{  {\fontsize{10pt}{12.0pt}\selectfont 300}} & 
		\multicolumn{1}{|P{0.47in}}{  {\fontsize{10pt}{12.0pt}\selectfont 1}} & 
		\multicolumn{1}{|P{0.47in}}{  {\fontsize{10pt}{12.0pt}\selectfont 0.008}} & 
		\multicolumn{1}{|P{0.47in}}{  {\fontsize{10pt}{12.0pt}\selectfont 25}} & 
		\multicolumn{1}{|P{0.47in}}{  {\fontsize{10pt}{12.0pt}\selectfont 20}} & 
		\multicolumn{1}{|P{0.47in}}{  {\fontsize{10pt}{12.0pt}\selectfont 150}} & 
		\multicolumn{1}{|P{0.23in}}{  {\fontsize{10pt}{12.0pt}\selectfont 11.5}} & 
		\multicolumn{1}{|P{1.01in}|}{  {\fontsize{10pt}{12.0pt}\selectfont 17.2}} \\
		\hline
		\multicolumn{1}{|P{0.47in}}{  {\fontsize{10pt}{12.0pt}\selectfont B4U}} & 
		\multicolumn{1}{|P{0.47in}}{  {\fontsize{10pt}{12.0pt}\selectfont 300}} & 
		\multicolumn{1}{|P{0.47in}}{  {\fontsize{10pt}{12.0pt}\selectfont 2}} & 
		\multicolumn{1}{|P{0.47in}}{  {\fontsize{10pt}{12.0pt}\selectfont 0}} & 
		\multicolumn{1}{|P{0.47in}}{  {\fontsize{10pt}{12.0pt}\selectfont 21}} & 
		\multicolumn{1}{|P{0.47in}}{  {\fontsize{10pt}{12.0pt}\selectfont 20}} & 
		\multicolumn{1}{|P{0.47in}}{  {\fontsize{10pt}{12.0pt}\selectfont 400}} & 
		\multicolumn{1}{|P{0.23in}}{  {\fontsize{10pt}{12.0pt}\selectfont 15}} & 
		\multicolumn{1}{|P{1.01in}|}{  {\fontsize{10pt}{12.0pt}\selectfont 14.4}} \\
		\hline
		\multicolumn{1}{|P{0.47in}}{  {\fontsize{10pt}{12.0pt}\selectfont B4D}} & 
		\multicolumn{1}{|P{0.47in}}{  {\fontsize{10pt}{12.0pt}\selectfont 300}} & 
		\multicolumn{1}{|P{0.47in}}{  {\fontsize{10pt}{12.0pt}\selectfont 2.5}} & 
		\multicolumn{1}{|P{0.47in}}{  {\fontsize{10pt}{12.0pt}\selectfont 0.39}} & 
		\multicolumn{1}{|P{0.47in}}{  {\fontsize{10pt}{12.0pt}\selectfont 13}} & 
		\multicolumn{1}{|P{0.47in}}{  {\fontsize{10pt}{12.0pt}\selectfont 20}} & 
		\multicolumn{1}{|P{0.47in}}{  {\fontsize{10pt}{12.0pt}\selectfont 400}} & 
		\multicolumn{1}{|P{0.23in}}{  {\fontsize{10pt}{12.0pt}\selectfont 10}} & 
		\multicolumn{1}{|P{1.01in}|}{  {\fontsize{10pt}{12.0pt}\selectfont 8.9}} \\
		\hline
		\multicolumn{1}{|P{0.47in}}{  {\fontsize{10pt}{12.0pt}\selectfont E3D}} & 
		\multicolumn{1}{|P{0.47in}}{  {\fontsize{10pt}{12.0pt}\selectfont 800}} & 
		\multicolumn{1}{|P{0.47in}}{  {\fontsize{10pt}{12.0pt}\selectfont 2.5}} & 
		\multicolumn{1}{|P{0.47in}}{  {\fontsize{10pt}{12.0pt}\selectfont 0.15}} & 
		\multicolumn{1}{|P{0.47in}}{  {\fontsize{10pt}{12.0pt}\selectfont 80}} & 
		\multicolumn{1}{|P{0.47in}}{  {\fontsize{10pt}{12.0pt}\selectfont 25}} & 
		\multicolumn{1}{|P{0.47in}}{  {\fontsize{10pt}{12.0pt}\selectfont 80}} & 
		\multicolumn{1}{|P{0.23in}}{  {\fontsize{10pt}{12.0pt}\selectfont 20}} & 
		\multicolumn{1}{|P{1.01in}|}{  {\fontsize{10pt}{12.0pt}\selectfont 16.5}} \\
		\hline
\end{tabular}
\caption{\label{Tab_one} List of the fitting parameters for the devices analyzed.}
\end{table}
%
%

\section{Simulation vs. experiment for the set of devices}\par

We present in the following the comparison between the experimental curves measured at 3.2 K (in blue) and the numerical simulation (in orange) based on the model detailed above.

%
%
\begin{table}[h]
 		\centering
		\begin{tabular}{p{2.95in}p{2.95in}}
		\hline
		\multicolumn{1}{|P{2.95in}}{
		\centering
		\includegraphics[width=0.35\textwidth]{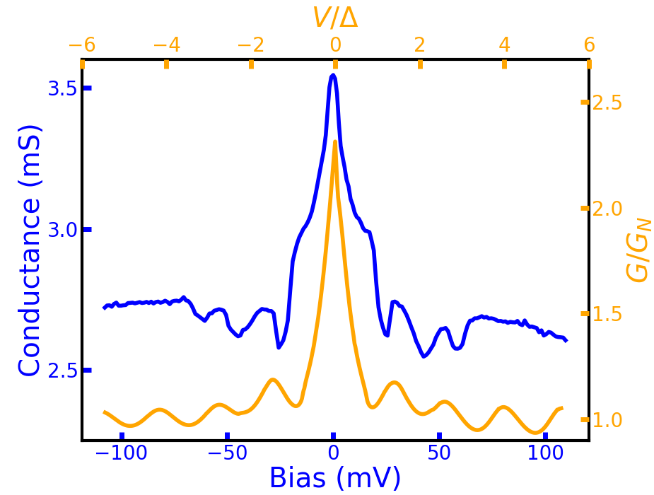}
	 	\par Device B4D} & 
		\multicolumn{1}{|P{2.95in}|}{
		\centering
		\includegraphics[width=0.35\textwidth]{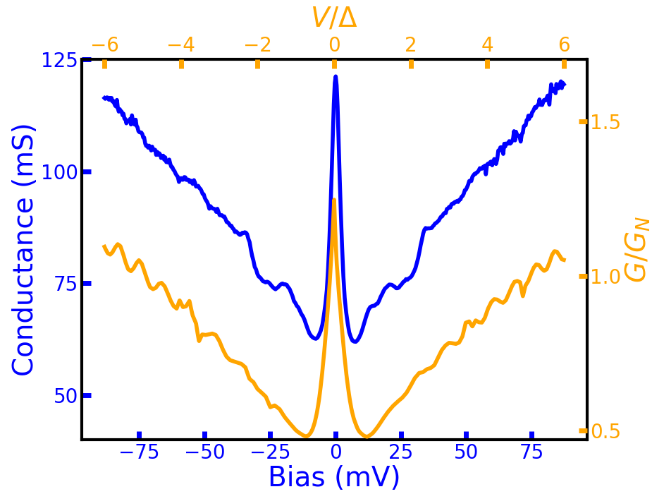}
		\par Device E3D} \\
		\hline
		\multicolumn{1}{|P{2.95in}}{
		\centering
		\includegraphics[width=0.35\textwidth]{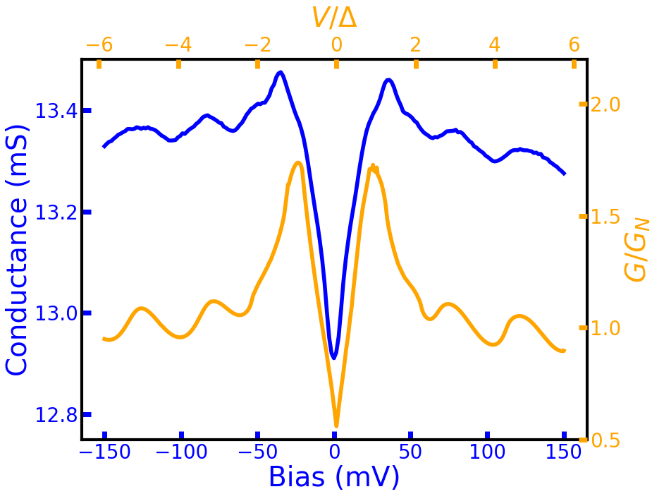}
		\par Device A3D} & 
		\multicolumn{1}{|P{2.95in}|}{
		\centering
		\includegraphics[width=0.35\textwidth]{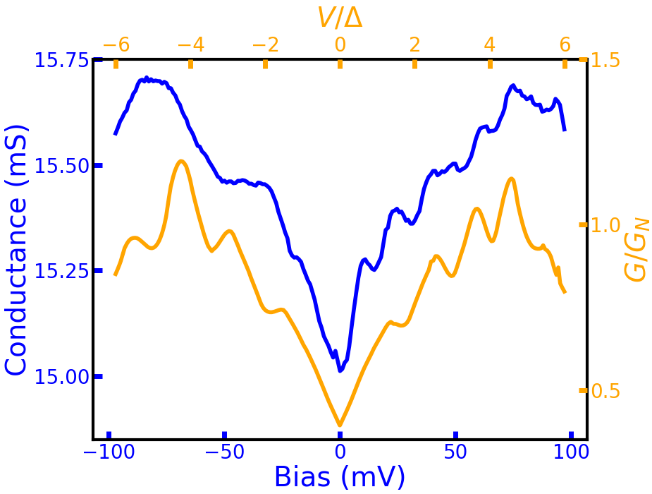}
		\par Device A3U} \\
		\hline
		\multicolumn{1}{|P{2.95in}}{
		\centering
		\includegraphics[width=0.35\textwidth]{image28.png}
		\par Device A5U} & 
		\multicolumn{1}{|P{2.95in}|}{
		\centering
		\includegraphics[width=0.35\textwidth]{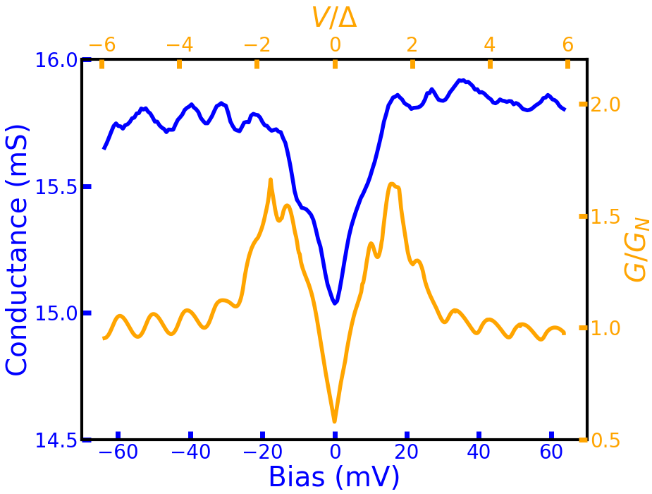}
		\par Device B4U} \\
		\hline
		\multicolumn{1}{|P{2.95in}}{
		\centering
		\includegraphics[width=0.35\textwidth]{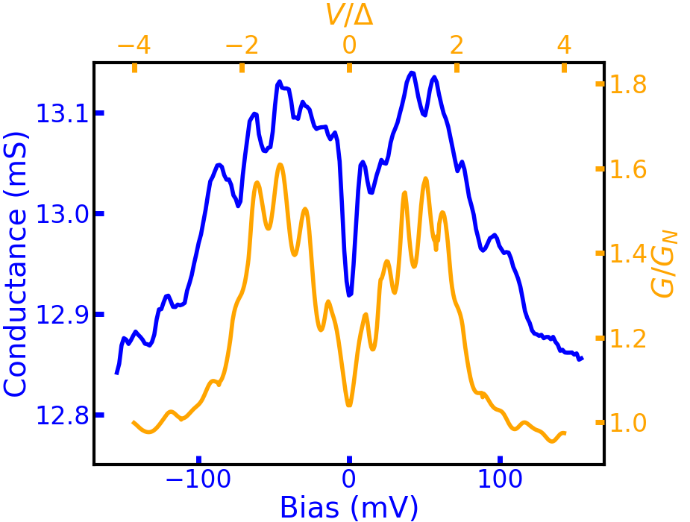}	
		\par Device B3U} & 
		\multicolumn{1}{|P{2.95in}|}{} \\
		\hline
\end{tabular}
\caption{\label{Tab_two} Comparison between the experimental curves measured at 3.2 K (in blue) and the numerical simulation (in orange) based on the model detailed above for all the devices analyzed.}
\end{table}
%
%

\clearpage
\section{Induced superconducting correlations: $d$-wave versus s-wave}

In principle and accordingly to~\cite{Balatsky_2006}, we do not expect the Au layer to significantly disturb the $d$-wave pairing, because it is one order-of-magnitude thinner than the mean free path in this material.

In our experiments, the key evidence for the propagation of the $d$-wave correlations into graphene is the observation of spectral features that cannot be explained based on s-wave superconductivity. In particular, the sharp zero-bias peak in some others [Fig. 3(a) in the main text and devices E3D and B4D] are characteristic of $d$-wave junctions cannot be reproduced by the theoretical calculations unless one considers $d$-wave correlations in the proximitzed graphene homojunction.

%
%
\begin{figure}[h]
	\centering
	\includegraphics[width=0.49\textwidth]{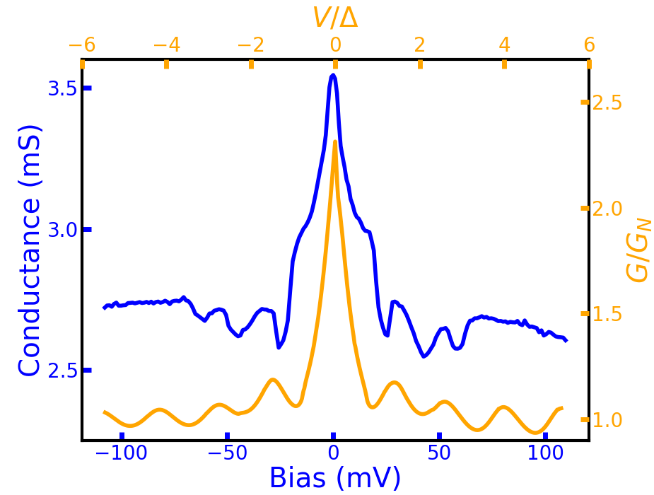}
	\caption{\label{Fig_twelve} Device \textbf{B4D}: Differential conductance experimental (blue) and theoretical (orange) as a function of the bias for $Z=2.5$, $\alpha=0.39$.}
\end{figure}
%
%

This is illustrated by the example shown in Fig.~\ref{Fig_twelve}, which corresponds to device B4D. The best agreement between theory and experiment is obtained with a barrier strength $Z=2.5$ and an effective angle between $d$-wave nodes and junction interfaces $\alpha=0.39$.

\begin{figure}[!h]
	\centering
	\begin{minipage}[c]{0.8\textwidth}
		\centering
		\includegraphics[width=0.39\textwidth]{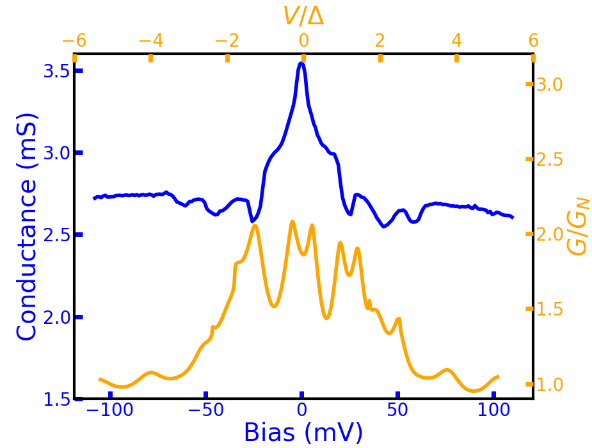}
		\includegraphics[width=0.39\textwidth]{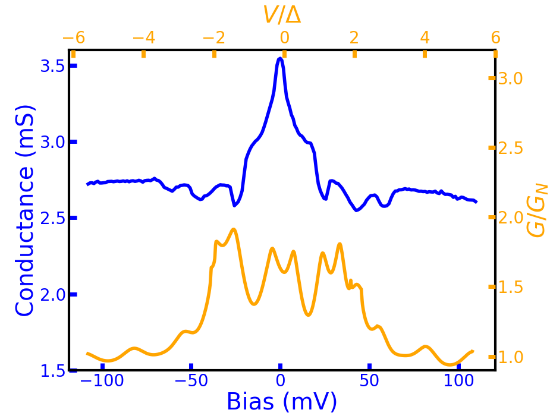}
	\end{minipage}
	\begin{minipage}[c]{0.8\textwidth}
		\centering
		\includegraphics[width=0.39\textwidth]{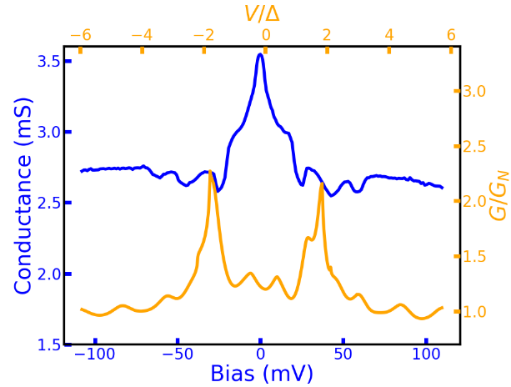}
		\includegraphics[width=0.39\textwidth]{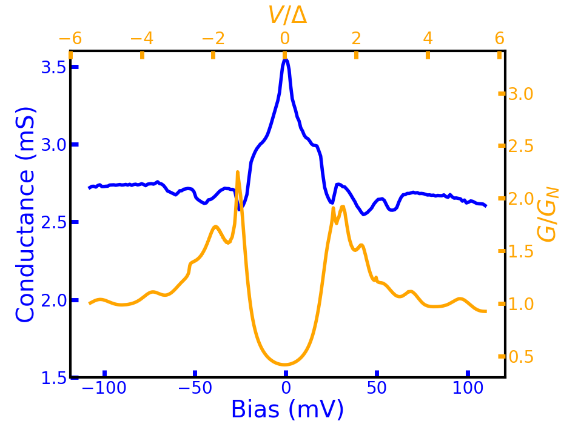}
	\end{minipage}
\caption{\label{Fig_thirteen} Differential conductance experimental (blue) and theoretical (0.39) as a function of the bias for different sets of fitting parameters for the device B4D assuming a $s$-wave proximization: $Z=1$ (top left), $Z=0.5$ (top right),   $Z=1$ (bottom left),  $Z=2$ (bottom right).}
\end{figure}

In Fig.~\ref{Fig_thirteen}, we show simulations based on $s$-wave superconductivity both at the YBCO/Au-graphene interface and across the proximitized graphene homojunction. One can see that this assumption does not allow reproducing the experimental behaviour, particularly the sharp zero-bias conductance peak, even if the barrier strength $Z$ is diminished to zero. Notice also that if the barrier strength $Z$ is finite, a ``dip"  emerges around zero bias where $s$-wave quasiparticle tunnelling is strongly supressed. This is just opposite to the behaviour observed in the experiments.

%
%
%
%

\begin{figure}[!th]
	\centering
	\begin{minipage}[c]{0.8\textwidth}
		\centering
		\includegraphics[width=0.39\textwidth]{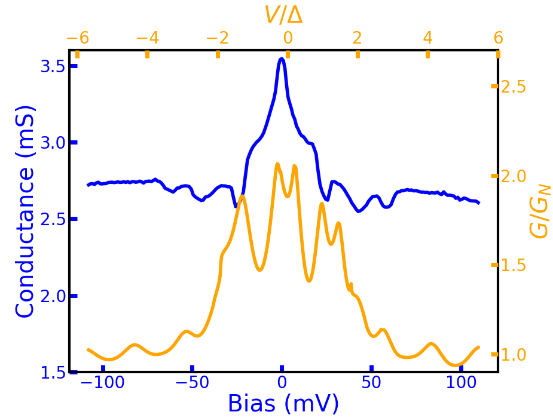}
		\includegraphics[width=0.39\textwidth]{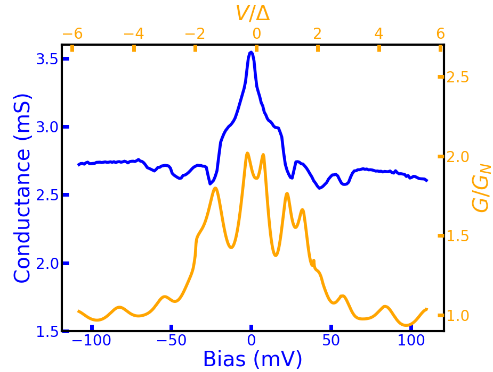}
	\end{minipage}
	\begin{minipage}[c]{0.8\textwidth}
		\centering
		\includegraphics[width=0.39\textwidth]{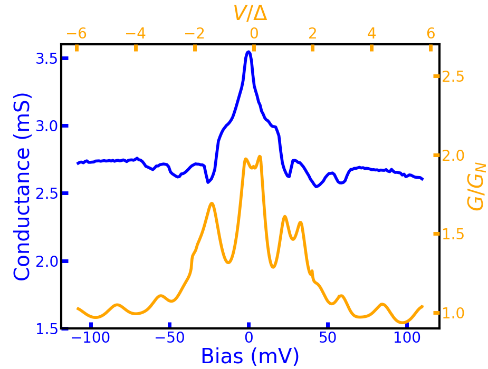}
		\includegraphics[width=0.39\textwidth]{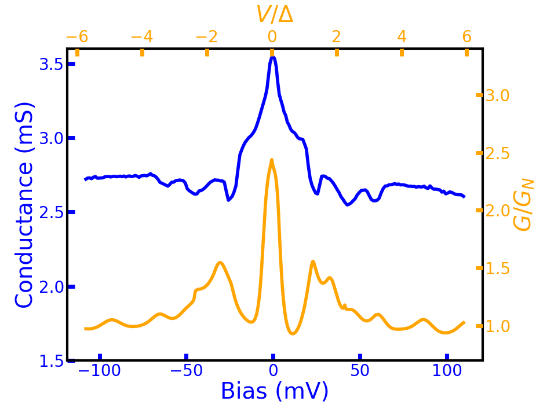}
	\end{minipage}
\caption{\label{Fig_fourteen} Differential conductance experimental (blue) and theoretical (0.39) as a function of the bias for different sets of fitting parameters for the device B4D assuming a $d$-wave proximization: $Z=1$ (top left), $Z=0.5$ (top right),   $Z=1$ (bottom left),  $Z=2$ (bottom right).}
\end{figure}

Finally, we show in Fig.~\ref{Fig_fourteen} simulations which consider $d$-wave superconductivity in the YBCO/Au-graphene interface and $s$-wave superconducting correlations across the graphene homojunction. One can see that under this assumption one cannot reproduce the behaviours observed experimentally. 
%
%
%
%

\begin{figure}[!h]
	\centering
	\begin{minipage}[c]{0.8\textwidth}
		\centering
		\includegraphics[width=0.39\textwidth]{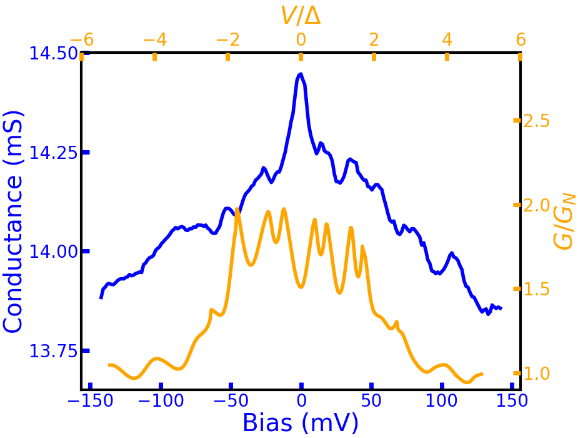}
		\includegraphics[width=0.39\textwidth]{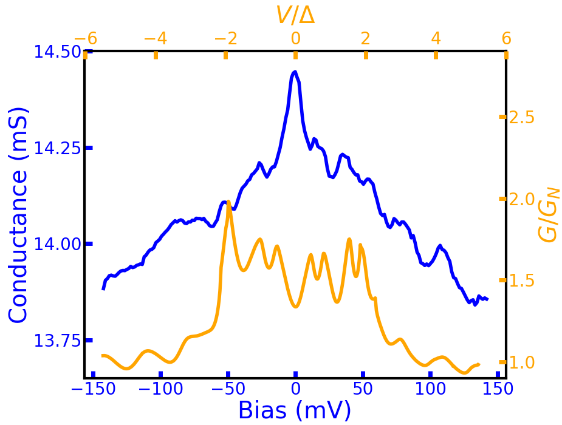}
	\end{minipage}
	\begin{minipage}[c]{0.8\textwidth}
		\centering
		\includegraphics[width=0.39\textwidth]{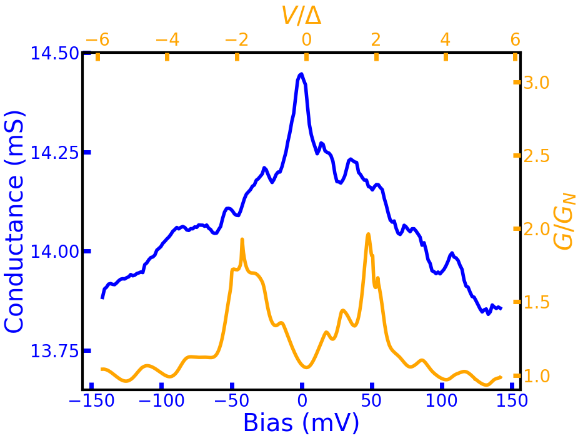}
		\includegraphics[width=0.39\textwidth]{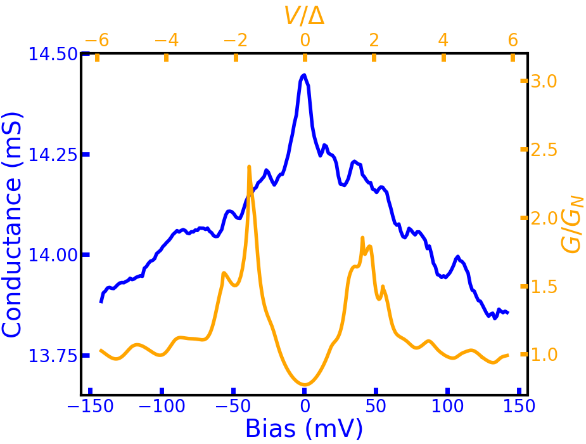}
	\end{minipage}
\caption{\label{Fig_fifteen} Differential conductance experimental (blue) and theoretical (0.39) as a function of the bias for different sets of fitting parameters for the device A5U assuming a $s$-wave proximization: $Z=1$ (top left), $Z=0.5$ (top right),   $Z=1$ (bottom left),  $Z=2$ (bottom right).}
\end{figure}

The next example corresponds to device A5U, results in Fig.~\ref{Fig_fifteen}. We assume $s$-wave correlations across every junction interface. Again, the zero-bias conductance peak cannot be reproduced by considering only $s$-wave correlations-wave propagating the junction interfaces.

%
%
%
%

\begin{figure}[!h]
	\centering
	\begin{minipage}[c]{0.8\textwidth}
		\centering
		\includegraphics[width=0.39\textwidth]{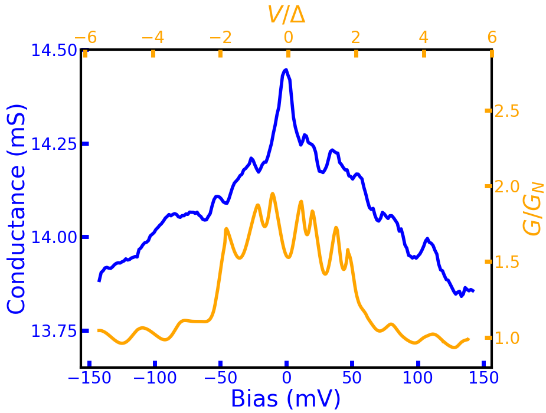}
		\includegraphics[width=0.39\textwidth]{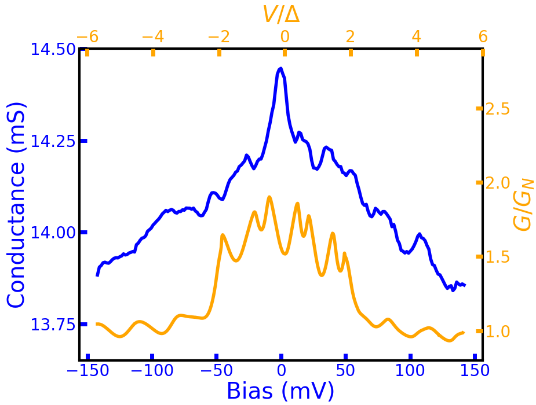}
	\end{minipage}
	\begin{minipage}[c]{0.8\textwidth}
		\centering
		\includegraphics[width=0.39\textwidth]{image61.png}
		\includegraphics[width=0.39\textwidth]{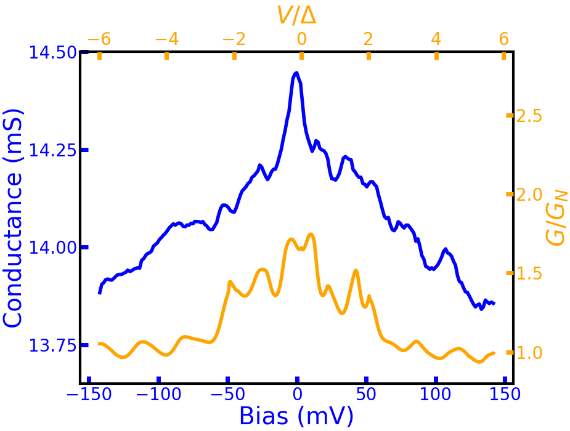}
	\end{minipage}
\caption{\label{Fig_sixteen} Differential conductance experimental (blue) and theoretical (0.39) as a function of the bias for different sets of fitting parameters for the device A5U assuming a $d$-wave proximization: $Z=1$ (top left), $Z=0.5$ (top right),   $Z=1$ (bottom left),  $Z=2$ (bottom right).}
\end{figure}

If we consider $d$-wave correlation only for the YBCO/Au-graphene interface and $s$-wave correlations across the homojunctions, the simulations do not allow reproducing the zero bias sharp zero bias conductance peak either, regardless of the barrier strength Z, see Fig.~\ref{Fig_sixteen}.

%
%
%
%

In summary, the experimental findings are well explained when it is assumed that $d$-wave superconducting correlations propagate into the graphene homojunction, and exclude the $s$-wave pairing component.

Notice finally that the effective angle  between the $d$-wave nodes and the junction interfaces changes randomly from device to device, which implies that it is not fixed by the device geometry. This suggests that the the contact between the graphene and the superconducting electrodes might be inhomoegeneous and occur at edges of the SC electrodes or at exposed $a$-axis facets in the YBCO surface~\cite{Sharoni_2004}. Local probes such as Scaning Tunneling microscopy would be necessary to investigate this question.

\bibliography{bibliography}